\documentclass[journal, twoside, web]{ieeecomm}

\usepackage{amsmath}
\usepackage{amssymb}
\usepackage{mathtools}
\usepackage{./TLC}
\usepackage{pgfplots}
\usepgfplotslibrary{groupplots}
\usetikzlibrary{spy}
\usepackage{multicol}
\usepackage{flushend}
\usepackage{graphicx}      
\usepackage[style=ieee,
            url=false,
            date=year]{biblatex}
\usepackage{multirow}
\usepackage{tasks}
\usepackage{cuted}
\usetikzlibrary{matrix,decorations.pathreplacing, calc, positioning,fit}

\def\BibTeX{{\rm B\kern-.05em{\sc i\kern-.025em b}\kern-.08em
    T\kern-.1667em\lower.7ex\hbox{E}\kern-.125emX}}

\title{Circuit realization and hardware linearization\\of monotone operator equilibrium networks}

\author{Thomas Chaffey \thanks{The author is with the School of Electrical and Computer Engineering, University of Sydney, New
South Wales, Australia. Email: {\tt\small thomas.chaffey@sydney.edu.au}.}%
}

\newcommand{\blue}[1]{\begingroup\color{black}#1\endgroup}

\begin{document}

\maketitle
\thispagestyle{empty}
\pagestyle{empty}

\begin{abstract}                
    It is shown that the port behavior of a resistor--diode network corresponds to
    the solution of a ReLU monotone operator equilibrium network (a neural
    network in the limit of infinite depth), giving a parsimonious construction of a
    neural network in analog hardware.  We furthermore show that the gradient of such a
    circuit can be computed directly in hardware, using a procedure we call
    \emph{hardware linearization}.  This allows the network to be trained in
    hardware, which we demonstrate with a device-level circuit simulation.
    We extend the results to cascades of resistor--diode networks, which can be used to
    implement feedforward and other asymmetric networks.  We finally show that
    different nonlinear elements give rise to different activation functions, and
    introduce the novel \emph{diode ReLU} which is induced by a 
    non-ideal diode model.
\end{abstract}

\section{Introduction}

The idea of building a neural network in analog hardware is classical
\cite{Hopfield1984, Hopfield1985, Koch1986, Harris1989a, Chua1988}.  Since the discovery of
semiconductor devices
with memristive properties \cite{Strukov2008}, and in light of the growing energy
intensiveness of machine learning systems, there has been a resurgence of interest in
building devices which incorporate analog memristive components and are specially
suited for deep learning applications \cite{Christensen2022a, Tye2023}.
\blue{Memristors and similar elements such as phase change memory have a resistance
    which can be varied electronically.  This allows the device to store a resistance
    value and act as a memory, as well as acting as a computational unit through
Ohm's law.} This allows the transport
delay between memory and computation to be
circumvented.  A particularly successful design is to arrange a number of memristors
in a crossbar array, which can be used to perform matrix-vector calculation
in a single operation \cite{Yao2020, Xia2019, Rasch2021, Wu2025a}.  Other recent designs seek to learn a
complete neural network mapping in analog hardware \cite{Kendall2020, Dillavou2023,
    Scellier2024a, Gao2025}.  Such designs are well suited to energy-based learning
    algorithms, which may be implemented directly in analog hardware,
    \blue{exploiting the reconfigurability of memristive devices for training and avoiding the
    energy overhead of a supervisory digital computer}
    \cite{Scellier2023, Huijzer2025, Dillavou2023, Scellier2017}.
It is unclear, however, what exactly is the class of neural networks that can be
learned in such hardware.

In this paper, we  begin by characterizing the class of neural networks that can be realized, 
for a restricted class of circuits: those
containing only LTI resistors, ideal transformers, ideal gyrators and ideal diodes
(RTGD networks).  It
is shown that these circuits correspond one-to-one with a class of machine learning
structures known as monotone operator equilibrium networks (monDEQs) \cite{Winston2020},
with ReLU activation
functions. These structures can be thought of as the limits of weight-tied neural networks in infinite depth.  This
theoretical correspondence gives a method of constructing a monDEQ using a small
number of passive devices, and performing a forward pass in a single operation,
by applying voltage or current excitations and reading the responses.  It also
provides a theoretical framework for studying existing analog machine learning
designs, such as the recently introduced Deep Resistive Networks \cite{Scellier2024},
which have been shown to be universal approximators of continuous functions
\cite{Scellier2024a}, and are a subclass of the circuits considered here.  Other
existing designs, such as Hopfield neural networks \cite{Hopfield1985} and
KirchhoffNets \cite{Gao2025}, incorporate dynamic elements and are not directly
captured by the circuits studied here, although it has been noted that
equilibrium models can be implemented as the equilibria of such dynamic circuits \cite{Wu2025}. The
advantage of using non-dynamic circuits as we do here is that there is no time of
convergence to the equilibrium; inference is (theoretically) instantaneous.

The correspondence between RTGD networks and monDEQs relies on the fact that both the monDEQ and circuit solve the
fixed-point of a particular monotone kernel equation.  In recent work, the author and
his colleagues have used monotone fixed-point methods to develop circuit simulation
algorithms \cite{Chaffey2023b, Chaffey2024, Shahhosseini2025}.  The connection
between electrical circuits and optimization algorithms is classical
\cite{Rockafellar1984}, but has
received renewed interest, in part due to the performance benefits possible with
analog computation \cite{Vichik2014,Achour2021, Boyd2024, Adegbege2024, Wu2025}.  The idea of studying deep neural
networks as monotone fixed-point problems originates in the work of
\cite{Combettes2020}.  There it is shown that, in the limit of infinite depth,
neural networks solve variational problems, which have also been studied in the
context of circuit theory \cite{Goeleven2017}.  monDEQs and related classes of
equilibrium models have been shown to perform well on standard machine learning
benchmarks, whilst requiring much less memory \cite{Winston2020, Bai2019}, and Lipschitz bounds can be enforced in
such models, guaranteeing robustness to perturbations \cite{Revay2020}.  

A further advantage of equilibrium networks is that their gradients can be computed by
implicitly differentiating the equilibrium condition, rather than backpropagating
layer by layer \cite{Winston2020, Bai2019}.  Our second main contribution is to show
that large parts of this implicit differentiation can be computed directly in
hardware, using a procedure we call \emph{hardware linearization}.  \blue{This procedure
exploits the reconfigurability of the memristive devices to train the device
directly, rather than train a model of the device and then realize the model in
hardware.} Hardware linearization involves
replacing the nonlinear elements in the circuit, in this case diodes, with their
linear approximations: either an open or short circuit.  For the special class of
\emph{reciprocal} circuits, the gradient computation simplifies further.

Reciprocity is a property originating in the theory of thermodynamics
\cite{Onsager1931}, which captures the fact that information flows symmetrically between the ports
of a circuit: information flows upstream as well as downstream.  The property was introduced to
systems theory by Willems \cite{Willems1972}; for a recent review, see 
\cite{vanderSchaft2025}.  In the context of learning, reciprocity allows
perturbations at the outputs of a circuit to be detected at the devices, and vice
versa -- it has been recognised as a necessary condition for some training 
methods, including equilibrium propagation for energy-based models
\cite{Scellier2017, Wanjura2025} and in-situ 
backpropagation for photonics \cite{Pai2023, Hughes2018}.  Here we show that the gradient of a
linear and reciprocal resistive network can be calculated by applying a perturbation voltage or current
at each device, and measuring the corresponding output.  

It has recently been
proposed to use gradient methods developed for machine learning to optimize analog
computers \cite{Achour2025}; here we develop a method to perform \emph{simultaneous}
learning and circuit optimization.  A bottleneck in the development of practical
analog machine learning architectures is dealing with the nonidealities of analog
devices \cite{Achour2021, Wu2025a}.  By computing gradients directly in hardware, hardware
linearization provides an avenue to automatically compensate for these nonidealities.
We perform a preliminary device-level circuit simulation which demonstrates this
advantage.


The remainder of this paper is structured as follows.  After introducing some
preliminary notation and definitions in Section~\ref{sec:prelims}, we establish the
neural network--circuit correspondence in two parts.  In Section~\ref{sec:circuit},
it is shown that the current--voltage behavior of a resistor--diode network can be
expressed as the kernel of the sum of two monotone operators.  In
Section~\ref{sec:mondeq}, it is shown that the same kernel equation corresponds to a
monDEQ, and we develop a detailed example based on a crossbar array structure.  This
structure is chosen due to its existing use in multipliers for deep learning.  We show that a crossbar array is able to realize a restricted class
of monDEQs, which in particular have symmetric weights matrices.
In Section~\ref{sec:linearization}, we introduce the method of
\emph{hardware linearization} to compute the gradient of a circuit directly in
hardware.
In
Section~\ref{sec:cascades}, we extend the results to cascades of networks, allowing
non-symmetric weights matrices to be implemented, and
show that feedforward ReLU networks may be implemented in this manner (recovering the
standard use of crossbar arrays as multipliers, as a special case of our theory).  In
Section~\ref{sec:activations}, we generalize to activations other than ReLUs.
We consider a non-ideal model of a diode and introduce a novel activation function, the
\emph{diode ReLU}.  We also show that saturation activations may be implemented using
pairs of ideal Zener and current regulator diodes.  Hardware linearization is demonstrated on a basic
circuit using a device-level circuit simulation in Section~\ref{sec:experiment}.
Conclusions and directions for future work are discussed in Section~\ref{sec:outlook}.

\section{Preliminaries}\label{sec:prelims}

Throughout this paper, $i \in \R$ refers to a current, and $v \in \R$ refers to a
voltage.  Vectors of currents and voltages will be denoted by $\mathbf{i}$ and
$\mathbf{v}$ respectively, while $\mathbf{u}$ and $\mathbf{y}$ refer to hybrid
vectors containing both currents and voltages. Subscripts ($i_k$ or $\mathbf{i}_k$)
refer either to port or driving point indices or to subcircuit indices, while vectors are indexed using square
brackets: the $k^\text{th}$ element of $\mathbf{i}$ is $\mathbf{i}[k]$.  Each voltage
$v$ across a port, driving point or device is naturally associated with the current
$i$ through the port, driving point or device, and the pair are referred to as
\emph{electrical duals}, with the terminology extending to vectors of voltages and
currents.     A (possibly
multi-valued) operator $R: \R^n \rightrightarrows \R^n$
between $\mathbf{u}$ and $\mathbf{y}$ is said to be $\mu$-\emph{strongly} \emph{monotone} if
$(\mathbf{u} - \mathbf{v})\cdot(\mathbf{y} - \mathbf{w}) \geq \mu \|\bu -
\mathbf{v}\|^2$ for some $\mu > 0$ for all $\mathbf{u},
\mathbf{v} \in \R^n$, $\mathbf{y} \in R(\mathbf{u})$, $\mathbf{w} \in R(\mathbf{v})$
and simply \emph{monotone} if $\mu = 0$. The operator $R$ is said to be \emph{maximal} if its
input--output graph $\operatorname{gph}R := \{ \mathbf{u}, \mathbf{y} \; | \; \mathbf{y} \in R(\mathbf{u})\}$
cannot be enlarged without violating monotonicity, and \emph{hybrid} if $\mathbf{u}$
and $\mathbf{y}$ are hybrid variables (containing a mix of currents and voltages).
Given an operator $\mathbf{v} = S(\mathbf{i})$  mapping a current to a voltage (that is, an impedance), its
electrical dual is the operator $\mathbf{i} = S(\mathbf{v})$, that is, the admittance with an identical
map. \blue{We use set-valued and single-valued notation interchangeably when a
    set-valued operator takes a singleton value, and make the standing assumption
that a set-valued operator takes the empty set outside its specified domain.}

\blue{Given a set $C \subseteq \R^n$ and a point $\mathbf{x} \in C$, the
\emph{tangent cone} to $C$ at $\mathbf{x}$, denoted $T_C(\mathbf{x})$, is
the set of all vectors $\mathbf{w} \in \R^n$ for which there exist sequences
$\mathbf{x}^\nu \to \mathbf{x}$ in $C$ and $\tau^\nu \searrow 0$ with
$(\mathbf{x}^\nu - \mathbf{x})/\tau^\nu \to
\mathbf{w}$~\autocite[Def.~6.1]{Rockafellar1998}.}

The $n\times d$ zero matrix is denoted $\mathbf{0}_{n \times d}$, or $\mathbf{0}$
when the dimension is clear from context or unimportant.  The column vector of ones
is denoted $\mathbf{1}$. \blue{A matrix $A \in \R^{n\times n}$ is said to be \emph{positive
semi-definite} if $x\tran A x \geq 0$ for all $x \in \R^n$, and \emph{positive
definite} if it is positive semi-definite and $x\tran A x = 0 \iff x = 0$. The linear
operator on $\R^n$ given by $x \mapsto Ax$ is maximal monotone if and only if $A$ is positive
semi-definite.}  A Stieltjes matrix is a symmetric, positive definite matrix
whose off-diagonal entries are non-positive.  Such a matrix $A$ is invertible and
$A^{-1}$ is entrywise non-negative (that is, $A$ is a nonsingular M-matrix). A
signature matrix is a diagonal matrix with diagonal entries of $\pm 1$.  Given
a vector $\mathbf{v}$, the diagonal matrix with $\mathbf{v}$ on the diagonal is denoted
$\operatorname{diag}(v)$.  Given matrices $A, B, \ldots$, the block-diagonal
matrix with $A, B, \ldots$ on the diagonal is denoted $\diag(A, B, \ldots)$, \blue{and the
block column matrix with blocks $A, B, \ldots$ is denoted $\col(A, B, \ldots)$.}

A graph is a collection of vertices and edges, $\mathcal{G} = (\mathcal{V},
\mathcal{E})$.  A tree is a graph which contains no cycles, and a spanning tree is a
subset of a graph which is a tree containing every vertex of the graph.  A Lagrangian
tree is a tree in which every edge shares a common vertex.

We consider non-dynamic circuits composed of linear, time-invariant (LTI)
resistors, ideal transformers, ideal gyrators and ideal diodes.  We refer to such
circuits as RTGD circuits. An $m$-\emph{port circuit} interacts with the environment via $m$
pairs of terminals, called \emph{driving points}, where the current in one terminal
is equal to the current out of the other.  We will study the \emph{driving point
behavior}, that is, the relation between the driving point currents and voltages.

The ideal elements are defined as follows.
An \emph{LTI resistor} defines a nonnegative linear relationship between a current $i \in \R$ and a
voltage $v \in \R$:  $v = R i$, $R \geq 0$.  An \emph{ideal} $(m+n)$-\emph{port transformer} is a
device with hybrid representation
\begin{IEEEeqnarray}{rCl}
    \begin{pmatrix} \mathbf{i}_1(t) \\ \mathbf{v}_2(t) \end{pmatrix} &=& \begin{pmatrix} 0 & T\tran \\ -T & 0 \end{pmatrix}\begin{pmatrix} \mathbf{v}_1(t) \\ \mathbf{i}_2(t) \end{pmatrix},
\end{IEEEeqnarray}
where $T \in \R^{n \times m}$ is the \emph{turns ratio matrix}, and each entry $T_{pq}$ of
$T$ gives the ratio of windings between driving point $p$ and driving point $q$.  An
\emph{ideal gyrator} is a 2-port device with impedance
\begin{IEEEeqnarray}{rCl}
    \begin{pmatrix} v_1(t) \\ v_2(t) \end{pmatrix} &=& \begin{pmatrix} 0 & \gamma \\ -\gamma & 0 \end{pmatrix}\begin{pmatrix} i_1(t) \\ i_2(t) \end{pmatrix},
\end{IEEEeqnarray}
where $\gamma \geq 0$ is the gyrator resistance.
The \emph{(forward biased) ideal diode} is defined as the circuit element with impedance
\begin{minipage}{0.5\linewidth}
\begin{IEEEeqnarray}{rCl}
v &\in& Z_{\text{diode}} (i) := \begin{cases} 0 & i > 0\\
        (-\infty, 0] & i = 0.
    \end{cases}\nonumber\\*
&&\nonumber
\end{IEEEeqnarray}
\end{minipage}%
\begin{minipage}{0.5\linewidth}
\centering
       \includegraphics{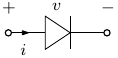}
\end{minipage}
\vspace{0.2cm}

\noindent The electrical dual of the forward biased ideal diode is the element with admittance
\begin{minipage}{0.5\linewidth}
\begin{IEEEeqnarray}{rCl}
i &\in& Y_{\text{diode}}(v) :=\begin{cases} 0 & v > 0\\
        (-\infty, 0] & v = 0,
    \end{cases}\nonumber\\*
&&\nonumber
\end{IEEEeqnarray}
\end{minipage}%
\begin{minipage}{0.5\linewidth}
\centering
\includegraphics{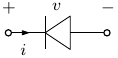}
\end{minipage}
\vspace{0.2cm}

\noindent which is a \emph{reverse biased} ideal diode.  We have used set-valued maps in these
definitions, so that both the forward and reverse biased
diodes define maximal monotone operators on $\R$.

\blue{We adopt standard notation for derivatives of single-valued mappings: for a
continuously differentiable function $f: \R^n \to \R^m$, we write $\nabla
f(\mathbf{x})
\in \R^{m \times n}$ for its Jacobian at $\mathbf{x}$, which we identify with the usual
gradient when $m = 1$.  For set-valued mappings, we use the \emph{graphical
derivative} of Rockafellar and Wets~\autocite[Def.~8.33]{Rockafellar1998}: given a
set-valued mapping $S: \R^n \rightrightarrows \R^m$ and a point $(\mathbf{x},
\mathbf{y}) \in
\operatorname{gph} S$, the graphical derivative $DS(\mathbf{x} \mid \mathbf{y}): \R^n
\rightrightarrows \R^m$ is the set-valued mapping whose graph is the tangent cone to
$\operatorname{gph} S$ at $(\mathbf{x}, \mathbf{y})$, that is,
\begin{IEEEeqnarray}{rCl}
w \in DS(\mathbf{x} \mid \mathbf{y})(u) & \iff & (u, w) \in T_{\operatorname{gph}
S}(\mathbf{x}, \mathbf{y}).
\end{IEEEeqnarray}
The graph of the forward biased ideal diode $Z_{\text{diode}}$ is the union of the
horizontal ray $\{(i, 0) : i > 0\}$ and the vertical ray $\{(0, v) : v \leq 0\}$.  A
direct calculation of tangent cones at any $(\bar \imath, \bar v) \in
\operatorname{gph} Z_{\text{diode}}$ gives
\begin{IEEEeqnarray}{rCl}
DZ_{\text{diode}}(\bar \imath \mid \bar v)(i) &=& \begin{cases}
    0 & \bar \imath > 0,\\
    Z_{\text{diode}}(i) & \bar \imath = 0,\ \bar v = 0,\\
    \R & \bar \imath = 0,\ \bar v < 0,\ i = 0.
\end{cases}
\end{IEEEeqnarray}
In particular, at the corner $(0, 0)$ the graphical derivative coincides with
$Z_{\text{diode}}$ itself.  A dual expression holds for the reverse biased diode
$Y_{\text{diode}}$, with the roles of $i$ and $v$ interchanged.}

Throughout the paper, we adopt the convention that a driving point current is
considered to be \emph{positive} if it is leaving the positive terminal (see
Figure~\ref{fig:RD_network}).

\blue{
The behavior of an RTG network may be described by a
\emph{hybrid matrix}, mapping a mix of currents and voltages to its electrical dual
vector. In some cases, the RTG network admits an \emph{impedance matrix},
mapping the driving point currents to the driving point voltages, and/or an \emph{admittance
matrix}, mapping the driving point voltages to the driving point currents. 
An RTG network described by a matrix $A$ (impedance, admittance or hybrid) is said to
be \emph{reciprocal} if there exists a signature matrix
$\Sigma$ such that $\Sigma A = A\tran \Sigma$ \cite{Willems1972}.  It is standard in
circuit theory to enforce that $\Sigma$ has $-1$ entries for each current and $+1$
for each voltage, in which case a circuit is reciprocal if it contains no gyrators
\cite{Anderson1973};
however, we do not enforce this additional constraint.}  

\begin{figure}[hb]
    \centering
    \includegraphics{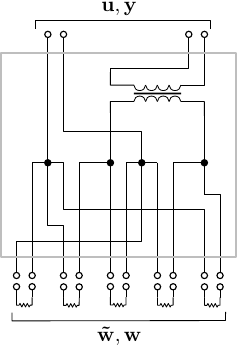}
    \caption{\blue{The element extraction approach to modelling an RTG circuit.}}
    \label{fig:element_extraction}
\end{figure}

\blue{We adopt the element extraction approach to circuit modelling \cite{Hughes2017,
    Anderson1973}.  Given an RTG network with resistances
    $\mathbf{r} \in \R_{\geq 0}^a$
    and conductances $\mathbf{g} \in \R_{\geq 0}^b$ for some $a, b \geq 0$, we can
    obtain a hybrid matrix description as follows.
    We begin by considering the circuit of wires, transformers and gyrators formed by
    replacing all resistors with internal ports, as illustrated in
Figure~\ref{fig:element_extraction}.} 

\blue{
    By solving Kirchhoff's current and voltage laws for this circuit, using  
    fundamental loop and cutset analysis \autocite{Seshu1961}, we can always find dual
    pairs of variables $\bu, \by$ (containing one current or voltage for each
    external port) and $\bw, \tilde
    \bw$ (likewise for resistors) and matrices $P, Q, M$ such that
    \begin{IEEEeqnarray}{rCl}
    \begin{pmatrix} \by \\ \tilde \bw  \end{pmatrix} &=& 
    \underbrace{\begin{pmatrix} P & -M\tran \\ M & Q \end{pmatrix}}_{F}
    \begin{pmatrix} \bu \\  \bw  \end{pmatrix}.\label{eq:WGT}
    \end{IEEEeqnarray}

    Without loss of generality, we assume $\bu$ and $\bw$ are ordered into currents
    followed by voltages. 
    Since this circuit is lossless, $F$, and in particular $P$ and $Q$, must be
    skew-symmetric \cite{Hughes2017}.  The
    resistors are described by
    \begin{IEEEeqnarray}{rCl}
    \tilde \bw &=& \diag(\mathbf{r}, \mathbf{g}) \bw.\label{eq:resistors}
    \end{IEEEeqnarray}
    Substituting 
    in \eqref{eq:WGT} and eliminating the variables $\tilde \bw, \bw$ gives
    \begin{IEEEeqnarray}{rCl}
    \by  &=& 
    \underbrace{\left(P - M\tran \left(\operatorname{diag}(\textbf{r}, \textbf{g}) -
    Q\right)^{-1} M\right)}_{\tilde H}
     \bu.
    \label{eq:tilde_H_partition}
    \end{IEEEeqnarray}
    We summarize this procedure in the following lemma, which also gives some
    properties of $\tilde H$.

\begin{lemma}\label{lem:RTG}
    Given an RTG network with resistances $\mathbf{r} \in \R_{\geq 0}^a$
    and conductances $\mathbf{g} \in \R_{\geq 0}^b$ for some $a, b \geq 0$, there exist dual
    pairs of variables $\bu, \by$  and $\bw, \tilde \bw$ and matrices $P, M, Q$ such
    that the network driving point behavior satisfies \eqref{eq:WGT} and
    \eqref{eq:tilde_H_partition}.  Furthermore:
    \begin{enumerate}
        \item $\tilde H + \tilde H\tran \preceq 0$;
        \item $\tilde H + \tilde H\tran \prec 0$ if and only if $M$ has full column
            rank, equivalently, for every
            $\mathbf{u} \neq 0$, at least one resistor or conductor carries a nonzero
            current/voltage in the resulting circuit response;
        \item if the circuit contains no gyrators, it is reciprocal: $\tilde H\tran =
            \Sigma \tilde H \Sigma$, where $\Sigma$ is a signature matrix with
            diagonal entry
            $-1$ if the corresponding entry of $\bu$ is a current, and diagonal entry
            $+1$ if it is a voltage.
    \end{enumerate}
\end{lemma}

\begin{proof}
    Existence of $\tilde H$, negative semi-definiteness 1) and the expression \eqref{eq:tilde_H_partition} are
    proven in \cite[Thm. 4]{Hughes2017}.  Reciprocity (part 3) follows from \cite[Thm.
    2.8.1]{Anderson1973}.  The full column rank condition in part 2 follows immediately from
    \eqref{eq:tilde_H_partition}.  From \eqref{eq:WGT}, it follows that $M$ is full
    column rank if and only if $\tilde \bw = 0 \iff \bu = 0$, that is, for every
            $\mathbf{u} \neq 0$, at least one resistor or conductor carries a nonzero
            current/voltage in the resulting circuit response.
\end{proof}
}

\section{From resistor--diode networks to kernel behaviors}\label{sec:circuit}

\begin{figure}[hb]
    \centering
    \includegraphics{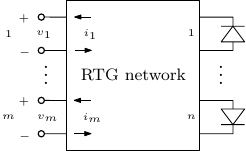}
    \caption{RTGD network with diodes extracted.}
    \label{fig:RD_network}
\end{figure}

Any $m$-port containing LTI resistors, transformers, gyrators and $n$ diodes can be drawn
in the form shown in Figure~\ref{fig:RD_network}, by dragging the diodes to the right
hand side.  To derive an expression for the port behavior of such a circuit, we then
extract the diodes, along the lines of classical element extraction procedures
\cite{Anderson1973, Hughes2017}, as follows:
\begin{enumerate}
    \item the diodes are cut off, producing an $(m+n)$-port RTG network and an $n$-port
    network containing $n$ disconnected diodes;
    \item the behavior of the RTG network is written as a hybrid matrix;
    \item the behavior of the diode network is written as a hybrid diagonal monotone
    operator;
    \item the networks are reconnected, giving a representation of the
    port behavior as the kernel of the sum of a positive semi-definite matrix and a diagonal
    monotone operator.
\end{enumerate}

Precisely, we have the following.

\begin{theorem}\label{thm:circuit}
        Let $\mathcal{C}$ be an $m$-port RTGD circuit, with $n$ diodes. Then there exist:
        \begin{enumerate}
                \item a partition of the driving point voltages $\{v_k\}_{k = 1\ldots
                m}$ and driving point currents $\{i_k\}_{k = 1\ldots m}$ into
                vectors $\mathbf{u}, \mathbf{y} \in \R^m$,
                \item vectors $\mathbf{i} \in \R^p$, $\mathbf{v} \in \R^q$, $p + q =
                n$, such that, for each diode, either the current through it is in $\mathbf{i}$ or
                the voltage across it is in $\mathbf{v}$,
                \item matrices $H \in \R^{n \times n}$, $H \succeq 0$, $B \in \R^{n\times m}$, $C \in \R^{m\times n}$, $D
                \in \R^{m\times m}$, $D \succeq 0$,
        \end{enumerate}
        such that
        \begin{IEEEeqnarray}{rCl}
                0 &\in& H \begin{pmatrix} \mathbf{i} \\ \mathbf{v} \end{pmatrix} +
                \begin{pmatrix} Z(\mathbf{i}) \\ Y(\mathbf{v}) \end{pmatrix} + B
                \mathbf{u},\IEEEyesnumber\label{eq:kernel}\\
                \mathbf{y} &=& -C \begin{pmatrix} \mathbf{i} \\ \mathbf{v}
                \end{pmatrix}- D \mathbf{u},\label{eq:output}
        \end{IEEEeqnarray}
        where the operators $Z$ and $Y$ are defined entrywise by $Z(\mathbf{i})[j]
        := Z_{\text{diode}} (\mathbf{i}[j])$ and $Y(\mathbf{v})[j]
        := Y_{\text{diode}} (\mathbf{v}[j])$, respectively.
\end{theorem}

\begin{proof}
    The proof follows the lines of \cite[Thm. 1]{Chaffey2023b}, but extracting only
    the diodes rather than every element.  \blue{Consider the circuit with diodes
        replaced by ports.  It then follows from Lemma~\ref{lem:RTG} that
        there exist $\tilde \bu, \tilde \by$ and $\tilde H$ such that the circuit
        satisfies \eqref{eq:tilde_H_partition} with inputs $\tilde \bu$ and outputs
        $\tilde \by$.  Partition $\tilde \bu = \col(\bu,
        \bz)$ and $\tilde \by = \col(\by, \tilde \bz)$, where $\bu, \by$
        correspond to external driving points and $\bz, \tilde \bz$ correspond to
        diode ports.  Reorder $\bz$ and $\tilde \bz$ so that
    $\mathbf{z} = 
    \col(\mathbf{i}, \mathbf{v})$, and $\tilde{\mathbf{z}}$ remains the
    electrical dual of $\bz$. 
    We then partition $\tilde H$ and pull out $-1$ so that 
    \begin{IEEEeqnarray}{rCl}
    \begin{pmatrix} \by \\ \tilde \bz \end{pmatrix} = \underbrace{-\begin{pmatrix} D & C \\ B & H
    \end{pmatrix}}_{\tilde H}\begin{pmatrix} \bu \\ \bz\end{pmatrix}. \label{eq:Hpartition}
\end{IEEEeqnarray}
    Substituting $\tilde z = \col(Z(\mathbf{i}), Y(\mathbf{v}))$
then yields the result.}
\end{proof}

Equation~\eqref{eq:kernel} is the kernel form of a behavior \cite{Willems1986} which
describes the internal currents and voltages of the circuit.
Theorem~\ref{thm:circuit} establishes one method of solving
Equations~\eqref{eq:kernel} and~\eqref{eq:output}: building the corresponding RTGD
circuit, applying the driving point inputs $\mathbf{u}$ and measuring the driving
point outputs
$\mathbf{y}$.  In the next section, we establish a second method of solving these
equations, using an optimization algorithm whose steps correspond to the layers of a
neural network.

\section{From kernel behaviors to equilibrium networks}\label{sec:mondeq}

The kernel equation~\eqref{eq:kernel} is in the form of a \emph{monotone inclusion}.
A large number of algorithms have been developed for solving such inclusions, in the
context of first-order convex optimization \cite{Ryu2022}.  One such algorithm is the
\emph{forward/backward algorithm}, \blue{which solves $0 \in F(\mathbf{x}) +
    G(\mathbf{x})$ for $\mathbf{x}$,
where $F$ and $G$ are maximal monotone operators, by the iteration
\begin{IEEEeqnarray}{rCl}
\mathbf{x}^{k+1} &=& (I + \alpha F)^{-1}(I - \alpha G)\mathbf{x}^k,
\end{IEEEeqnarray}
where $\alpha > 0$ and the superscripts denote iteration count.  If $G$ is strongly
monotone, then there exists a range for $\alpha$ such that $(I - \alpha G)$ is a
contraction \cite{Ryu2022} and the iteration converges to a unique fixed point by Banach's fixed
point theorem.  Setting $\bz := \col(\mathbf{i}, \mathbf{v})$, $\psi := \col(Z(\mathbf{i}),
Y(\mathbf{v}))$, $F(\bz) := \psi(\bz)$ and $G\bz := H\bz + B\bu$ gives the iteration}
\begin{IEEEeqnarray}{rCl}
     \begin{pmatrix} \mathbf{i}^{k+1} \\ \mathbf{v}^{k+1} \end{pmatrix} &=& (I +
     \alpha \psi)^{-1}\left((I - \alpha H)\begin{pmatrix} \mathbf{i}^k \\
     \mathbf{v}^k \end{pmatrix} - \blue{\alpha} B\bu\right).
\label{eq:fb}
\end{IEEEeqnarray}

It was shown in \cite{Combettes2020}, and in the context of circuit theory in 
\cite{Chaffey2023b}, that $(I + \alpha \psi)^{-1}(\cdot) = \relu{\cdot}$,
where
\begin{IEEEeqnarray}{rCl}
    \relu{u} &:=& \begin{cases}
        u & u \geq 0\\
        0 & \text{otherwise.}
    \end{cases}
\end{IEEEeqnarray}
Equation~\eqref{eq:fb} therefore represents one layer of a ReLU neural network, with
weights matrix $(I - \alpha H)$, and the fixed-point solution to \eqref{eq:fb}
represents the output of a weight-tied neural network in the limit of infinite depth.
Such limits have been studied in the machine learning literature under the name
\emph{equilibrium models} \cite{Bai2019}.  The particular case here, when $H \succeq
0$ and $\psi$ is given by a collection of ideal diodes,
gives rise to a  \emph{ReLU monotone operator 
equilibrium network} \cite{Winston2020}, which we define formally as follows.  

\begin{definition}
 Given maximal monotone $\psi: \R^n \to \R^n$ and positive semidefinite $H$, suppose
there exists a fixed point $\col(\mathbf{i}^\star,\; \mathbf{v}^\star)$ for the iteration
\eqref{eq:fb}.  Then the mapping from $\bu$ to $\col(\mathbf{i}^\star,\;
\mathbf{v}^\star)$
is called a \emph{monotone operator equilibrium network} (monDEQ).
\end{definition}

\blue{The fixed point $\col(\mathbf{i}^\star,\; \mathbf{v}^\star)$ is guaranteed to
    exist when the forward/backward iteration $\eqref{eq:fb}$ is a contraction.  A
sufficient condition is that $\tilde H \prec 0$ as given by Lemma~\ref{lem:RTG} part
2, in which case $H \succ 0$. All the examples in this paper satisfy this condition.}
Note that, in general, many optimization algorithms may be applied to solve
\eqref{eq:kernel}, each inducing an equivalent monDEQ with the same set of fixed
points but possibly different convergence properties (see \cite{Winston2020} for
details).  For this reason, we consider the circuit kernel equations
\eqref{eq:kernel}--\eqref{eq:output} to be the fundamental object.  

Applying the forward/backward algorithm to \eqref{eq:kernel} establishes that each
RTGD network realizes a monDEQ.
By invoking a converse result from classical circuit theory \cite{Anderson1973}, we 
establish the following
correspondence as a corollary to Theorem~\ref{thm:circuit}.

\begin{corollary}\label{cor:correspondence}
    The kernel \eqref{eq:kernel} of an RTGD network is the set of fixed-points of a
    ReLU monotone operator equilibrium network, and given a ReLU monotone operator equilibrium network,
    such an RTGD network always exists.
\end{corollary}

\begin{proof}
    Substitution of a fixed point $\col(\mathbf{i}^{k+1},\; \mathbf{v}^{k+1}) =
    \col(\mathbf{i}^{k},\; \mathbf{v}^{k})$
    in \eqref{eq:fb} immediately yields \eqref{eq:kernel}, establishing one direction
    of the correspondence.
Furthermore, given a positive semi-definite matrix $H$, there always
exists a network of resistors, transformers and gyrators with impedance matrix $H$
\cite[$\S$9.2, p. 373]{Anderson1973}, which establishes the other direction.
\end{proof}

\subsection{Example 1: Crossbar arrays}\label{sec:example1}

\begin{figure}[hb]
    \centering
    \includegraphics{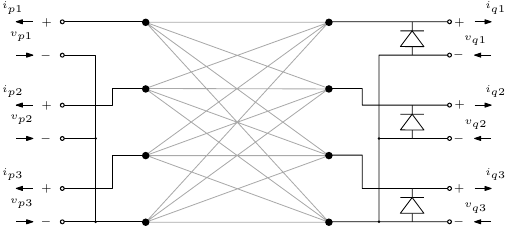}
    \caption{A crossbar array.  Grey edges are resistors.  At input ports (left), voltages are
    treated as inputs and currents as outputs.  At output ports (right), currents are treated as
inputs and voltages as outputs.}
    \label{fig:crossbar1}
\end{figure}

A crossbar array is a network of resistors 
with a fully connected, bipartite interconnection graph, between a set of input nodes
and output nodes.  Such arrays are widely studied as accelerators for matrix-vector
computations \cite{Xia2019, Heidema2024}, where they are used to implement a
\emph{single layer} of a feedforward neural network. In this section, we show that
crossbar arrays can also be used to implement an equilibrium layer when accessed
using a particular port configuration, illustrated in 
Figure~\ref{fig:crossbar1}.  It turns out that this structure is only able to
represent a restricted class of monDEQs, which motivates the construction of a richer
class of networks using these circuits as basic building blocks, which will be
developed in Section~\ref{sec:cascades}. 
We let $p-1$ denote the number of input ports and $q-1$ the
number of output ports, which is equal to the number of diodes.  The number of
resistors is then equal to $pq$.

\begin{figure}[hb]
    \centering
    \includegraphics{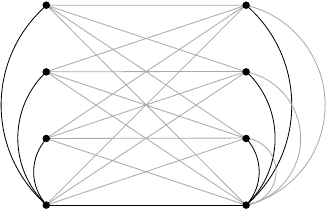}
    \caption{The circuit graph of the circuit in Figure~\ref{fig:crossbar1}, with a
    spanning tree in black, containing input ports, diodes and a single resistive edge.}
    \label{fig:crossbar_tree}
\end{figure}

In order to derive the kernel behavior \eqref{eq:kernel} of this circuit, we first
partition the circuit graph into a spanning tree and its complement, the cotree,
such that
\begin{enumerate}
    \item any edge whose voltage we wish to treat as an input belongs to the tree;
    \item any edge whose current we wish to treat as an input belongs to the cotree.
\end{enumerate}
Such a partition is illustrated in Figure~\ref{fig:crossbar_tree}, for the circuit of
Figure~\ref{fig:crossbar1}.

We then apply fundamental cut set and loop analysis \cite{Seshu1961} to this tree/cotree partition to
express the cotree voltages as a function of the tree voltages, and the tree currents
as a function of the cotree currents.  This gives a matrix $F$ in \eqref{eq:WGT} with the following structure, where blank spaces
have zero entries and $-N\tran$ is the $(p+q) \times (pq-1)$ incidence matrix of the subgraph containing
the resistive edges in the cotree:

\begin{equation}
\begin{pmatrix} \mathbf{v}_q \\ \mathbf{i}_p \\
\mathbf{i}_\text{diode} \\ \mathbf{v}_R \\ i_R \end{pmatrix} =
\begin{tikzpicture}[scale=0.9,baseline=(current bounding box.center)]
    \centering
    \draw[line width=0.6pt] (0,6) -- (6,6) -- (6,0) -- (0,0) -- cycle;

    \draw[line width=1.0pt] (3,0) -- (3,6);
    \draw[line width=0.6pt] (1,1) -- (1,4);
    \draw[line width=0.6pt] (2,5) -- (2,6);
    \draw[line width=0.6pt] (5,0) -- (5,5);

    \draw[line width=1.0pt] (0,3) -- (6,3);
    \draw[line width=0.6pt] (1,1) -- (6,1);
    \draw[line width=0.6pt] (0,4) -- (1,4);
    \draw[line width=0.6pt] (2,5) -- (5,5);

    \node at (0.5,3.5) {$-I$};
    \node at (2.5,5.5) {$I_{q\times q}$};
\node at (4.0,4.0) {$-N\tran$};
    \node at (5.5,2.0) {$\mathbf{1}_{pq-1}$};
    \node at (2.0,2.0) {$\mathit{N}$};
    \node at (4,0.5) {$-\mathbf{1}^T$};
\end{tikzpicture}
\begin{pmatrix} \mathbf{i}_q \\ \mathbf{v}_p \\
\mathbf{v}_\text{diode} \\ \mathbf{i}_R \\ v_R \end{pmatrix}.
\label{eq:crossbar_F}
\end{equation}

We furthermore have the following
characterization of $D$ and $H$ in the kernel representation \eqref{eq:kernel}--\eqref{eq:output}.

\begin{proposition}\label{prop:Stieltjes}
    Let $\tilde H$ as defined in Lemma~\ref{lem:RTG} correspond to a circuit of the form of Figure~\ref{fig:crossbar1}
    and assume that no resistive edge is open or short.
    Partition $\tilde H$ according to \eqref{eq:Hpartition}, and further partition
    $D$ according to
    \begin{equation}
  D=
\begin{tikzpicture}[
  baseline,
  label distance=10pt 
]

\matrix [matrix of math nodes,left delimiter=(,right delimiter=),row sep=0.1cm,column sep=0.1cm] (m) {
D_{11} & D_{12} \\ D_{21} & D_{22}\\ };

\node[
  fit=(m-1-1)(m-1-1),
  inner xsep=0,
  above delimiter=\{,
  label=above:$q$
] {};

\node[
  fit=(m-1-2)(m-1-2),
  inner xsep=0,
  above delimiter=\{,
  label=above:$p$
 ] {};

\node[
  fit=(m-1-2)(m-1-2),
  inner xsep=15pt,inner ysep=0,
  right delimiter=\},
  label=right:$q$
] {};

\node[
  fit=(m-2-2)(m-2-2),
  inner xsep=15pt,inner ysep=0,
  right delimiter=\},
  label=right:$p.$
] {};

\end{tikzpicture}
\end{equation}
Then $D_{11} = \mathbf{0}$ and $D_{22}$ and $H$ are Stieltjes matrices.
\end{proposition}

\begin{proof}
    \blue{The circuit contains no gyrators, so } it follows from Lemma~\ref{lem:RTG}
    part 3 that the matrix $\tilde H$ is signature
    symmetric \blue{with respect to the signature matrix}
\begin{IEEEeqnarray*}{rCl}
\Sigma = \begin{pmatrix} -I_{q\times q} & 0 \\ 0 & I_{(p+q)\times(p+q)} \end{pmatrix},
\end{IEEEeqnarray*}
so we have $\Sigma \tilde H  = \tilde H\tran \Sigma$.
    Note that $N$ is the fundamental circuit matrix of the subcircuit containing
    all $(p+q+2)$ nodes and the $pq - 1$ resistive edges in the cotree.  It
    follows from \cite[$\S$4-2]{Seshu1961} that $N$ has full row rank, \blue{so $M$
    has rank $(pq-1)$}. \blue{It also follows from the expression for $F$ that $X =
    (\diag(\mathbf{r}, \mathbf{g}) - Q)^{-1}$ has full rank}. It then follows that
    $\tilde H$ has the structure
    \begin{IEEEeqnarray}{rCl}
        \tilde H &=& \begin{pmatrix} \mathbf{0}_{q \times q} & \mathbf{0}_{q \times
        (p+q)} \\ \mathbf{0}_{(p+q) \times q} & N\tran X N
    \end{pmatrix}.\label{eq:tilde_H_form}
    \end{IEEEeqnarray}
    It follows that $D_{11} = \mathbf{0}$ and
    $D_{22}$ and $H$ are full rank.

    We now consider the subcircuit formed by shorting input ports (setting
    $\mathbf{v}_\text{in} = 0$), replacing output ports
    with open circuits (setting $\mathbf{i}_\text{out} = 0$) and replacing diodes with ports.  The 
    resulting circuit has admittance matrix $H$ 
    \blue{mapping $\mathbf{v}_\text{diode}$ to $\mathbf{i}_\text{diode}$}.
    Its port graph is a Lagrangian tree with every edge
    oriented towards the common node, \blue{hence $H$ is a \emph{nodal} impedance
    matrix and therefore a weighted Laplacian. In particular,} $H$ is diagonally dominant with nonpositive off-diagonal entries and positive
    diagonal entries \cite[Lemma 4.1]{Chen2019},
    \cite[Thm. 3]{Brown1961}, \cite{Guillemin1960}.
      
    Since $H$ is full rank, it is positive definite.  Finally, signature
    symmetry of $\tilde H$ implies symmetry of $H$.  It follows that $H$ is a
    Stieltjes matrix.  
    That $D_{22}$ is Stieltjes is proved in an analogous manner,
    but shorting the diodes (setting $\mathbf{v}_\text{diode} = 0$) and replacing the
    output ports with open circuits (setting $\mathbf{i}_\text{out} = 0$).  
\end{proof}

The fact that $H$ and $D_{22}$ are Stieltjes matrices restricts the class
of maps that may be learned.  It will be investigated in Section~\ref{sec:cascades}
how interconnecting multiple crossbar arrays can be used to alleviate this issue.
Stieltjes matrices, and the more general class of M-matrices, have a long history in
the study of electrical networks \cite{Stieltjes1887}, and have attractive computational
properties.  For example, it is known that Successive Over-Relaxation (SOR)
converges for Stieltjes matrices with an easily calculated near-optimal relaxation factor \cite[$\S$12.2, Thm 2.1]{Young1971}.  SOR can
be used to calculate the circuit gradient using equilibrium backpropagation
\cite{Winston2020}, however experimentally it does not perform as well as
Douglas-Rachford with the optimal step size of \cite{Giselsson2017}.  In the following
section, we will show that the gradient can also be computed directly in hardware.


\section{Hardware linearization of equilibrium networks}\label{sec:linearization}

One of the main advantages of equilibrium networks is that the gradient of the
network, with respect to the parameters $\theta$, can be calculated by
differentiating the  equilibrium condition directly, rather than backpropagating
through multiple layers.  In this section, we show that this gradient calculation can
also be performed directly in hardware, using a fixed, small number of operations.

We consider a kernel behavior whose matrices and input depend \blue{continuously} on a set of parameters,
collected in a vector $\theta \in \R^r$:
\begin{IEEEeqnarray}{rCl}
0 &\in& H_\theta \mathbf{z} +
\psi(\mathbf{z}) + B_\theta
        \mathbf{u}_\theta,\IEEEyesnumber\label{eq:kernel_param}\\
\mathbf{y} &=& -C_\theta \mathbf{z} - D_\theta
\mathbf{u}_\theta.\label{eq:output_param}
\end{IEEEeqnarray}
\blue{Throughout this section, we make the standing assumption that $H_\theta \succ 0$ for
all $\theta \in \R^r$.}
We define a cost $\mathcal{C}(\mathbf{y})$ which we would like to minimize over the parameters
$\theta$.  Taking the derivative of $\mathcal{C}(\mathbf{y})$ with respect to $\theta$ gives
\begin{IEEEeqnarray*}{rCl}
\blue{\nabla_\theta \mathcal{C}} &=& \blue{\nabla_{\by} \mathcal{C}}\left(\blue{\nabla_\theta \by}\right)\\
\blue{\nabla_\theta \by} &=& -C_\theta
\blue{\nabla_\theta \bz}  - D_\theta \blue{\nabla_\theta \bu_\theta}\\
&& {}- \blue{\nabla_\theta C_\theta}\,\bz -
\blue{\nabla_\theta D_\theta}\,\bu_\theta.\IEEEyesnumber\label{eq:lin_output}
\end{IEEEeqnarray*}

\blue{Assuming that no diode is at its switching point\footnotemark, we can apply
    Lemma~\ref{lem:implicit_diff}
    to implicitly differentiate \eqref{eq:kernel_param} at the operating point
$(\bz, \boldsymbol{\eta})$, where $\boldsymbol{\eta} := -H_\theta \bz - B_\theta
\bu_\theta \in \psi(\bz)$, which} gives
\begin{IEEEeqnarray*}{rCl}
0 &\in& H_\theta \blue{\nabla_\theta \bz} + \blue{D\psi(\bz \mid
\boldsymbol{\eta})\,\nabla_\theta \bz} \\&&+ B_\theta
\blue{\nabla_\theta \bu_\theta} + \blue{\nabla_\theta H_\theta}\, \bz + \blue{\nabla_\theta B_\theta}\,
\bu_\theta.\IEEEyesnumber\label{eq:lin_fixed_point}
\end{IEEEeqnarray*}


\footnotetext{
    \blue{
    If a diode is at the switching point $(v = 0, i = 0)$, an analogous result to
    Lemma~\ref{lem:implicit_diff} holds for the directional derivative rather than
    the gradient.  However, for any practical implementation, the diode will not be
    ideal and will not have a sharp switching point, and even for the ideal diode,
    the switching point is a set of measure zero in the graph of the operator, so we
do not develop the directional derivative case here.}
}

It is shown in \cite{Winston2020} that \eqref{eq:lin_fixed_point} can be solved using
a splitting algorithm such as forward/backward, similarly to inference using a
monDEQ.  
The remainder of this section establishes that large parts of the computation of
\blue{$\nabla_\theta \bz$} and \blue{$\nabla_\theta \by$} can be computed
directly in hardware.  We do this in two steps.  Firstly, we show that the implicit
equation \eqref{eq:lin_fixed_point} can be solved
by hardware-linearizing the original RTGD network with kernel behavior
\eqref{eq:kernel_param}--\eqref{eq:output_param}.  Secondly, we show that when the
hardware linearized network is reciprocal, the gradient of the linearized circuit may
be computed by applying a voltage or current over each resistor.

Hardware linearization is
formalized below in Theorem~\ref{thm:hardware_linearization}, and involves replacing
the diodes of the original network with their hardware-linearized equivalents, which
we now define.

\begin{definition}
    Given an operating point \blue{$(\bar \imath, \bar v) \in \operatorname{gph} Z_{\text{diode}}$}
    and an offset $v_d \in \R$, the \emph{hardware
    linearization at \blue{$(\bar \imath, \bar v)$} with offset $v_d$} of the forward-biased ideal diode is the element mapping $i_\ell \in \R$ to
    $v_\ell \in \R$ according to
    \begin{IEEEeqnarray*}{rCl}
    v_\ell &\in& \blue{DZ_{\text{diode}}(\bar \imath \mid \bar v)(i_\ell) + v_d}
         \\ &=& \blue{\begin{cases}
            v_d & \bar \imath > 0,\\
            Z_{\text{diode}}(i_\ell) + v_d & \bar \imath = 0,\ \bar v = 0,\\
            \R & \bar \imath = 0,\ \bar v < 0,\ i_\ell = 0.
        \end{cases}}
    \end{IEEEeqnarray*}
    \begin{minipage}{\linewidth}
    \centering
           \includegraphics{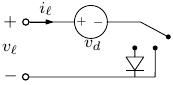}
       \end{minipage}
       \vspace{0.2cm}

    Given an operating point \blue{$(\bar v, \bar \imath) \in \operatorname{gph} Y_{\text{diode}}$}
    and an offset $i_d \in \R$, the \emph{hardware
    linearization at \blue{$(\bar v, \bar \imath)$} with offset $i_d$} of the reverse-biased ideal diode is the element mapping $v_\ell \in \R$ to
    $i_\ell \in \R$ according to
    \begin{IEEEeqnarray*}{rCl}
    i_\ell &\in& \blue{DY_{\text{diode}}(\bar v \mid \bar \imath)(v_\ell) + i_d}\\
           &=& \blue{\begin{cases}
            i_d & \bar v > 0,\\
            Y_{\text{diode}}(v_\ell) + i_d & \bar v = 0,\ \bar \imath = 0,\\
            \R & \bar v = 0,\ \bar \imath < 0,\ v_\ell = 0.
        \end{cases}}
    \end{IEEEeqnarray*}
    \begin{minipage}{\linewidth}
    \centering
           \includegraphics{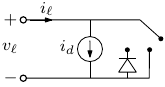}
       \end{minipage}
       \vspace{0.2cm}
\end{definition}

\begin{theorem}\label{thm:hardware_linearization}
    Given an RTGD network with kernel behavior
    \eqref{eq:kernel_param}--\eqref{eq:output_param} at a particular operating point
    $(\bu_\theta, \bz, \by)$\blue{, with $\boldsymbol{\eta} := -H_\theta \bz - B_\theta \bu_\theta \in \psi(\bz)$}, the RTG network formed by replacing the diodes with their
    hardware linearizations at \blue{$(\bz, \boldsymbol{\eta})$} with offsets $\bu_d$ has the kernel behavior
    \begin{IEEEeqnarray}{rCl}
    0 &\in& H_\theta \bz_\ell + \blue{D\psi(\bz \mid \boldsymbol{\eta})}\,\bz_\ell + B_\theta \mathbf{u}_\ell + \bu_d, \IEEEyesnumber\label{eq:kernel_lin}\\
    \by_\ell &=& -C_\theta \bz_\ell - D_\theta \bu_\ell.\label{eq:output_lin}
    \end{IEEEeqnarray}
\end{theorem}
\begin{proof}
    The proof follows directly from substituting the hardware linearization for each
    entry of $\psi(\bz)$ in the kernel behavior
    \eqref{eq:kernel_param}--\eqref{eq:output_param}.
\end{proof}

Theorem~\ref{thm:hardware_linearization} allows us to compute the gradient
\blue{$\nabla_\theta \by$} in hardware, as follows.

\begin{corollary}\label{cor:gradient_computation}
    Given an RTGD network \eqref{eq:kernel_param}--\eqref{eq:output_param} at a particular operating point
    $(\bu_\theta, \bz, \by)$ where no diodes are at their switching point, the gradient \blue{$\nabla_\theta \by$} can be
    computed as follows:
    \begin{enumerate}
        \item Hardware linearize the circuit at $\bz$ with offset and input
            \begin{IEEEeqnarray*}{rCl}
            \bu_d &=& \blue{\nabla_\theta H_\theta}\, \bz + \blue{\nabla_\theta B_\theta}\,
\bu_\theta \\
            \bu_\ell &=& \blue{\nabla_\theta \bu_\theta},\IEEEyesnumber\label{eq:grad_offset}
            \end{IEEEeqnarray*}
        \item Compute \blue{$\nabla_\theta \by$} as
            \begin{IEEEeqnarray*}{rCl}
            \blue{\nabla_\theta \by} &=& \by_\ell - \blue{\nabla_\theta C_\theta}\,\bz -
\blue{\nabla_\theta D_\theta}\, \bu_\theta.\IEEEyesnumber\label{eq:grad_output}
            \end{IEEEeqnarray*}
    \end{enumerate}
\end{corollary}
\begin{proof}
    Substituting the given $\bu_d$ and $\bu_\ell$ in the kernel behavior
    \eqref{eq:kernel_lin}--\eqref{eq:output_lin} and adding
    the specified terms to the output $\by_\ell$ gives equations \eqref{eq:lin_output} and
    \eqref{eq:lin_fixed_point}, with \blue{$\bz_\ell = \nabla_\theta \bz$}.
\end{proof}

Corollary~\ref{cor:gradient_computation} allows the implicit equation
\eqref{eq:lin_fixed_point} to be solved in hardware, in a single step.  However, the
matrix derivatives \blue{$\nabla_\theta H,
\nabla_\theta B, \nabla_\theta C$} and
\blue{$\nabla_\theta D$} have to be precomputed.  The following theorem shows
that, in the special case that the resistive network is reciprocal and $\theta$
contains the resistances and conductances, these derivatives can also
be computed in hardware.

\begin{theorem}\label{thm:reciprocal_backprop}
    Consider a hardware linearized reciprocal RTD circuit with $a$ resistors, $b$ conductors and a hybrid matrix
    $\tilde H$ as defined in
    \eqref{eq:tilde_H_partition}.  Let $\theta = (\mathbf{r}, \mathbf{g})\tran \in \R^{a + b}$, $\Sigma = \diag(-I_a, I_b)$ and fix $i \in \{1,
    2, \ldots, a + b\}$.  Measure hybrid vectors $\by, \tilde{\bz}$ as follows:
    \begin{enumerate}
        \item For each diode, open the linearization circuit if the diode is
            forward-biased, and close the circuit if the diode is reverse-biased.
        \item If edge $i$ is a resistor, apply a unit voltage across the edge.  If
            edge $i$ is a conductor, apply a unit current through the edge.
        \item Read the vector of outputs $\by$ and the vector of diode responses
            $\tilde{\bz}$ as defined in \eqref{eq:WGT}.
    \end{enumerate}
    \begin{IEEEeqnarray}{lrCl}
    \text{Then }\quad & \pd{\tilde H}{\theta_i} &=& \Sigma_{ii} \begin{pmatrix} \by
        \\ \tilde{\bz} \end{pmatrix} \begin{pmatrix} \by\tran & \tilde{\bz}\tran
    \end{pmatrix}\begin{pmatrix} \Sigma_{\by} & \\ &
        \Sigma_{\tilde{\bz}} \end{pmatrix},\label{eq:recip_backprop}
    \end{IEEEeqnarray}
    where $\Sigma_{\by}$ and $\Sigma_{\tilde{\bz}}$ are signature matrices whose diagonal
    entries are $-1$ if the corresponding entries in $\by$ (respectively
    $\tilde{\bz}$) are currents, and are $1$ otherwise, and $\Sigma_{ii}$ denotes the
    $i^{\text{th}}$ diagonal entry of $\Sigma$.
\end{theorem}

\begin{proof}
    Let $\Theta := \diag(\theta)$.
    Differentiating \eqref{eq:tilde_H_partition} with respect to $\theta_i$ and
    applying Lemma~\ref{lem:inverse_derivative} in the appendix gives
    \begin{IEEEeqnarray*}{rCl}
    \pd{\tilde H}{\theta_i} &=& -\pd{}{\theta_i} M\tran\left(\Theta - Q\right)^{-1}M\\
        &=& -M\tran \pd{}{\theta_i} \left( \Theta - Q\right)^{-1}M\\
        &=& M\tran(\Theta-Q)^{-1}\pd{}{\theta_i}(\Theta -
        Q)(\Theta - Q)^{-1} M\\
        &=& M\tran(\Theta-Q)^{-1}\ei\ei\tran(\Theta - Q)^{-1} M,
        \IEEEyesnumber\label{eq:intermediate_1}
    \end{IEEEeqnarray*}
    where $\ei$ is the $i^{\text{th}}$ canonical basis vector.

 Note that the circuit is reciprocal and $\theta$ is partitioned into resistances
 followed by conductances, so $\tilde H$ is signature symmetric with signature
    matrix $\Sigma = \diag(-I_a, I_b)$.
    Proceeding with the element extraction approach used in the proof of
    Lemma~\ref{lem:RTG}, setting external
    inputs \blue{$\bu$} to zero (shorting ports with voltage inputs, opening ports with current
    inputs) \blue{and neglecting external outputs $\by$}, \blue{Equation~\eqref{eq:WGT} simplifies to
    \begin{IEEEeqnarray}{rCl}
    \tilde \bw = Q \bw.\label{eq:recip_pf_1}
\end{IEEEeqnarray}}
    Applying a
    voltage (current) across (through) each resistor (conductor) \blue{in the vector
    $\mathbf{o}$} gives a device law of
    \begin{IEEEeqnarray}{rCl}
    \tilde \bw &=& \diag(\mathbf{r}, \mathbf{g}) \bw - \mathbf{o}.\label{eq:recip_pf_2}
    \end{IEEEeqnarray}
    Combining \blue{\eqref{eq:recip_pf_1} and \eqref{eq:recip_pf_2}} gives the kernel behavior
    \begin{IEEEeqnarray}{rCl}
    \diag(\mathbf{r}, \mathbf{g})\bw - Q\bw - \mathbf{o} = 0, \label{eq:internal}
    \end{IEEEeqnarray}
    Setting $\mathbf{o} = \ei$ in
    \eqref{eq:internal} gives 
    $\bw_1 = (\Theta-Q)^{-1}\ei.$
    Setting $\mathbf{o} = \Sigma \ei = \Sigma_{ii} \ei$ (which is negation if
    $\mathbf{o}[i]$ is a voltage, and
    identity otherwise) gives 
    $\bw_2 = (\Theta-Q)^{-1}\Sigma\ei,$
    so 
    \begin{IEEEeqnarray*}{rCl}
    \bw_2\tran\Sigma &=& \ei\tran \Sigma((\Theta-Q)^{-1})\tran \Sigma\\
                   &=& \ei\tran (\Theta-Q)^{-1},
    \end{IEEEeqnarray*}
    where the final equality follows from reciprocity.  The identity
    \begin{IEEEeqnarray*}{rCl}
    \pd{\tilde H}{\theta_i} &=& \Sigma_{ii} M\tran \bw_1 \bw_1\tran
    \Sigma M
    \end{IEEEeqnarray*}
    then follows from \eqref{eq:intermediate_1}, noting that, by linearity, $\bw_2 =
    \Sigma_{ii} \bw_1$.  From \eqref{eq:WGT}, we have that
    \begin{IEEEeqnarray}{rCl}
    \begin{pmatrix} \by \\ \tilde{\bz} \end{pmatrix} &=& P\begin{pmatrix} \bu \\ \bz
\end{pmatrix} - M\tran \bw.\label{eq:original_identity}
    \end{IEEEeqnarray}
    Step 1 in the theorem statement guarantees $\bz = 0$, and $\bu = 0$ as there are
    no external excitations.  Therefore we have
    \begin{IEEEeqnarray}{rCl}
    \begin{pmatrix} \by \\ \tilde{\bz} \end{pmatrix} &=& -M\tran
    \bw.\label{eq:one_side}
    \end{IEEEeqnarray}
    Furthermore, 
    \begin{IEEEeqnarray}{rCl}
    -M\tran \Sigma \bw &=& \begin{pmatrix} \Sigma_{\by} & \\ &
        \Sigma_{\tilde{\bz}} \end{pmatrix} \begin{pmatrix} \by \\ \tilde{\bz}
    \end{pmatrix}.\label{eq:other_side}
    \end{IEEEeqnarray}
    The result then follows by substituting \eqref{eq:one_side} and
    \eqref{eq:other_side} in \eqref{eq:original_identity}.
\end{proof}

The method of Theorem~\ref{thm:reciprocal_backprop} is closely related to the method
of \emph{in-situ backpropagation} in photonic neural networks, which also exploits
reciprocity \cite{Pai2023,
Hughes2018}.  There, however, the adjustable parameters are permittivities of
phase-shifters rather than resistances, and this difference results in a different
test procedure.  Theorem~\ref{thm:reciprocal_backprop} is also related to the method of
adjoint sensitivity analysis, which has been used in circuit design optimization by
simulating a second adjoint circuit \cite{Joshi2017}.

The method of hardware linearization presented in this section calculates the
gradient of the circuit output with respect to a parameter \emph{precisely}.  This
differs from other methods for training analog circuits in hardware, such as
Contrastive Learning, which computes the gradient of the contrastive cost function
\cite{Movellan1991, Huijzer2025} and Equilibrium Propagation, which approximates the
gradient of an arbitrary cost in the limit of small nudging parameter
\cite{Scellier2017}.  \blue{We assume that each diode in the circuit may be switched.
    Such a capability could be realized by pairing each diode with a simple CMOS
    circuit, in a manner similar to the 1T1R Resistive RAM configuration
    \cite{Hsu2004}.  The design of a such a circuit is a topic for future research.}

\section{Cascade networks}\label{sec:cascades}

\begin{figure}[hbp]
    \centering
    \includegraphics[width=\linewidth]{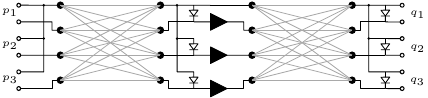}
    \caption{Cascade of two crossbar arrays, each configured according to
    Figure~\ref{fig:crossbar1}.  The black triangles represent ideal amplifiers,
whose gains are collected in the vector $\sigma$ of synaptic weights.}
    \label{fig:cascade}
\end{figure}

The property that $\tilde H$ in the previous example is signature symmetric is a
general property, known as \emph{reciprocity}.  Any RTGD network which contains no
gyrators will have a signature symmetric hybrid matrix $\tilde H$ (and hence 
submatrix $H$)
\cite{Anderson1973}.  This is a significant restriction on the equilibrium networks
which may be realized: in practice, it is often desirable to have a lower triangular
matrix $H$, which can be used to represent convolutional structures
\cite{Winston2020}.  One remedy is
to incorporate gyrators in the design.  While the ideal passive gyrator has not been
realized (without an external field), it can be realized using active components such as operational amplifiers.
In this section, we pursue a more direct method of obtaining a lower triangular $H$,
using a cascade of subcircuits (Figure~\ref{fig:cascade}).  The subcircuits are connected by ideal amplifiers
with gains $\sigma$. In analogy with biological neural networks, we refer to these
gains as \emph{synaptic weights}.  We note that such cascades may be realized
physically, by building a cascade of several circuits, or with a single
reconfigurable array which is used to compute the output of each layer in sequence.
This latter approach is the standard method for accelerating neural network inference
using a crossbar array \cite{Xia2019}.  The networks introduced in this section
include as a special case Deep Resistive Networks \cite{Scellier2024},
which have been shown to be universal approximators of continuous functions
\cite{Scellier2024a}.  We show in Section~\ref{sec:example2} that they also include
feedforward ReLU neural networks, which are also known to be universal approximators
\autocite{Leshno1993, Yarotsky2017}.

The following proposition gives the kernel
behavior resulting from a cascade of RTGD networks.  The inputs are split into two vectors, allowing only some of the outputs
of the previous subcircuit to be cascaded (for example, only voltages and not
currents, as in Figure~\ref{fig:cascade}).

\begin{proposition}
    Let $(d_i)_{i = 1, 2, \ldots, \ell}$ be a sequence of positive integers.
    Let $(\mathcal{C}_i)_{i = 1, 2, \ldots, \ell}$ be a sequence of RTGD circuits with
    kernel equations
    \begin{IEEEeqnarray*}{rCl}
    0 &\in& H_i \mathbf{z}_i + \psi_i(\mathbf{z}_i)
             +
            B_i
            \mathbf{u}_i + \tilde{B}_i \tilde{\bu}_i,\\
    \mathbf{y}_i &=& -C_i \mathbf{z}_i - D_i \mathbf{u}_i - \tilde{D}_i\tilde{\bu}_i,\IEEEyesnumber\label{eq:cascade_component}
    \end{IEEEeqnarray*}
    where $\mathbf{z}_i := (\mathbf{i}_i\tran  \mathbf{v}_i\tran)\tran$,
    $\psi_i(\mathbf{z}_i) := (Z_i(\mathbf{i}_i)\tran Y_i(\mathbf{v}_i)\tran )\tran$,
    for each $i = 2, \ldots, \ell$, $\mathbf{u}_{i}, \mathbf{y}_{i-1} \in \R^{d_i}$, and $\mathbf{u}_{i}$ and $\mathbf{y}_{i-1}$ are compatible in
    the sense that if, given $k$, $\mathbf{u}_{i}[k]$ is a  voltage (respectively, current), so
    is  $\mathbf{y}_{i-1}[k]$.  Let $\sigma_i \in \R^{d_i}$ for each $i$ and let
    $\boldsymbol{\sigma}_i$ denote $\diag(\sigma_i)$. Then, the circuit formed by setting
    $\mathbf{u}_{i} = \boldsymbol{\sigma}_i \mathbf{y}_{i-1}$ for all $i$ is described 
    by the kernel equation
    \begin{IEEEeqnarray}{rCl}
    \label{eq:cascade}
    0 &\in& H \mathbf{z} +
    \psi(\mathbf{z}) + B
    \mathbf{u}_1 + \tilde{B}\tilde{\bu},\IEEEyessubnumber\label{eq:cascade_kernel}\\
    \mathbf{y} &=& -C \mathbf{z} - D
    \mathbf{u}_1 - \tilde{D}\tilde{\bu},\IEEEyessubnumber\label{eq:cascade_output}
    \end{IEEEeqnarray}
    where
    {\footnotesize
    \begin{IEEEeqnarray*}{rCl}
    \mathbf{z} &:=& (\mathbf{z}_1\tran \mathbf{z}_2\tran \ldots
    \mathbf{z}_\ell\tran)\tran,\\
    \psi(\bz) &:=& (\psi_1(\mathbf{z}_1)\tran,
    \psi_2(\mathbf{z}_2)\tran \ldots \psi_\ell(\mathbf{z}_\ell)\tran)\tran\\
    \tilde{\bu} &:=& (\tilde{\bu}_1\tran \tilde{\bu}_2\tran \ldots
    \tilde{\bu}_\ell\tran)\tran,\\
    H &:=& \left(\begin{array}{l l l l l} H_1 & 0 & 0 & 0 & \ldots\\
            -B_2\boldsymbol{\sigma}_2 C_1 & H_2 & 0 & 0 & \ldots \\
            B_3\boldsymbol{\sigma}_3 D_2 C_1 & -B_3\boldsymbol{\sigma}_3 C_2 & H_3 & 0 & \ldots\\
            -B_4\boldsymbol{\sigma}_4 D_3 D_2 C_1 & B_4\boldsymbol{\sigma}_4 D_3 C_2
                                                  & -B_4\boldsymbol{\sigma}_4 C_3 & H_4 & \ldots\\
                        \vdots & \vdots & \vdots & \vdots & \ddots
               \end{array}\right)\\
            B &:=& \left(\begin{array}{l} B_1\boldsymbol{\sigma}_1 \\
                    -B_2\boldsymbol{\sigma}_2 D_1 \\
            B_3\boldsymbol{\sigma}_3 D_2
            D_1 \\ -B_4\boldsymbol{\sigma}_4 D_3 D_2 D_1 \\
            \vdots
        \end{array}\right)\\
            \tilde{B} &:=& \left(\begin{array}{l l l l l} \tilde{B}_1 & 0 & 0 & 0 & \ldots \\ 
                    -B_2\tilde{D}_1 & \tilde{B}_2 & 0 & 0 & \ldots \\ 
                    B_3 D_2 \tilde{D}_1 & -B_3\tilde{D}_2 & \tilde{B}_3 & 0 & \ldots \\ 
                    -B_4 D_3 D_2 \tilde{D}_1 & B_4 D_3 \tilde{D}_2 & -B_4 \tilde{D}_3
                                             & \tilde{B}_4 & \ldots\\
                    \vdots & \vdots & \vdots & \vdots & \ddots
        \end{array}\right)\\
                    C &:=& \blue{\left(\ldots \; -D_\ell D_{\ell-1} D_{\ell-2} C_{\ell-3},
                    \; D_\ell D_{\ell-1} C_{\ell-2},\; -D_{\ell} C_{\ell-1}, \;
            C_\ell \right)}\\
                        D &:=& \blue{(-1)^{\ell+1}D_\ell D_{\ell-1} \ldots D_1}\\
                        \tilde{D} &:=& \blue{\left(\ldots \; -D_\ell D_{\ell-1} D_{\ell-2}
                            \tilde{D}_{\ell-3}, \; D_\ell D_{\ell-1}
                            \tilde{D}_{\ell-2},\;
                    -D_\ell\tilde{D}_{\ell-1}, \; \tilde{D}_\ell \right)}.\nonumber\\*
                    &&\IEEEyesnumber\label{eq:cascade_matrices}
            \end{IEEEeqnarray*}}
\end{proposition}

\begin{proof}
    The proof follows from a straightforward calculation, iteratively substituting
    the expression for $\mathbf{y}_{i - 1}$ in the kernel equations for
    $\mathcal{C}_i$.
\end{proof}

\subsection{Hardware linearization of cascade networks}\label{sec:linearization_cascade}

We now address the question of how to hardware linearize a cascade of RTGD networks.
It turns out that gradients flow through cascaded networks in quite a natural way.
To compute the gradient of a downstream output with respect to a parameter in an
upstream circuit, we use hardware linearization to compute the gradient of the
upstream circuit as in Section~\ref{sec:linearization}, then apply the upstream
gradient as input to the hardware linearized downstream circuit and measure its
output.  This procedure is formalized in the following theorem. 

\begin{theorem}\label{thm:cascade_linearization}
    Consider an RTGD circuit $\mathcal{C}$, described by the kernel behavior
    \begin{IEEEeqnarray}{rCl}
    0 &\in& H \mathbf{z} + \psi(\mathbf{z}) + B
    \mathbf{u}(\theta),\label{eq:lincas_kernel}\\
    \mathbf{y} &=& -C \mathbf{z} - D
    \mathbf{u}(\theta),\label{eq:lincas_output}
    \end{IEEEeqnarray}
where $\theta\in\R$ is a parameter.  Then, given
    $\td{\mathbf{u}}{\theta}$, the derivative $\td{\mathbf{y}}{\theta}$ may be
    computed as follows:
    \begin{enumerate}
        \item hardware linearize $\mathcal{C}$ at the input $\mathbf{u}$;
        \item apply $\td{\mathbf{u}}{\theta}$ as input to the hardware linearized
            $\mathcal{C}$.  Then the output of the hardware linearized
            $\mathcal{C}$ is $\td{\mathbf{y}}{\theta}$.
    \end{enumerate}
\end{theorem}
\begin{proof}
    Differentiating \eqref{eq:lincas_kernel}--\eqref{eq:lincas_output} with respect
    to $\theta$ \blue{at the operating point $(\bz, \boldsymbol{\eta})$, where
    $\boldsymbol{\eta} := -H \bz - B \bu(\theta) \in \psi(\bz)$,} gives
    \begin{IEEEeqnarray*}{rCl}
    0 &\in& H \td{\mathbf{z}}{\theta} +
    \blue{D\psi(\bz \mid \boldsymbol{\eta})}\,\td{\bz}{\theta} + B
    \td{\mathbf{u}}{\theta},\\
    \td{\mathbf{y}}{\theta} &=& -C \td{\mathbf{z}}{\theta} - D
    \td{\mathbf{u}}{\theta},
    \end{IEEEeqnarray*}
    which is the kernel equation of $\mathcal{C}$ hardware linearized at the input
    $\bu$ with input $\td{\bu}{\theta}$.
\end{proof}

\subsection{Example 2: feedforward networks}\label{sec:example2}

\begin{figure}[ht]
    \centering
    \includegraphics{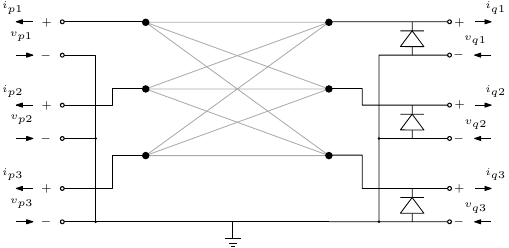}
    \caption{A crossbar array configured with a common earth terminal for all ports
        and diodes.  Grey edges are resistors.  At $p$ ports, voltages are
    treated as inputs and currents as outputs.  At $q$ ports, currents are treated as
inputs and voltages as outputs.}
    \label{fig:crossbar2}
\end{figure}

We now consider a slightly different configuration of a crossbar array which, when
cascaded, allows us to build a purely feedforward network.  This configuration is
illustrated in Figure~\ref{fig:crossbar2}.  Under this configuration, the matrix $H$
becomes diagonal, and the kernel equation can be solved explicitly and reduces to the
application of a ReLU.  The circuit therefore implements a single layer of a
feedforward ReLU network.  By duplicating inputs and cascading several layers with
positive and negative synaptic weights, we can implement an arbitrary feedforward
ReLU neural network, recovering the standard use of crossbar arrays in deep learning
(see, for instance, \autocite{Yao2020}) as a special case of the theory presented in this paper.

\begin{strip}
\begin{IEEEeqnarray}{rCl}
\scriptsize
    \left(\begin{array}{c | c c c | c c c}
            \mathbf{0} & \multicolumn{3}{c}{\mathbf{0}} & \multicolumn{3}{c}{I}\\\hline
            \multirow{3}{*}{$\mathbf{0}$} & g_{11} + g_{12} + \ldots & 0 & \ldots & -g_{11} & -g_{12} & \ldots\\
                                          & 0 & g_{21} + g_{22} + \ldots & \ldots & -g_{21} & -g_{22} & \ldots\\ 
                                          & \vdots & \vdots & \ddots & \vdots & \vdots & \ddots\\\hline
            \multirow{3}{*}{$I$} & -g_{11} & -g_{21} & \ldots  & g_{11} + g_{21} +
            \ldots & 0 &\ldots \\
                                 & -g_{12} & -g_{22} & \ldots  & 0 & g_{12} + g_{22}
                                + \ldots&\ldots\\ 
                                 & \vdots & \ddots   & \vdots & \vdots & \ddots & \vdots 
                        \end{array}\right)\label{eq:feedforward_kernel_0}
\end{IEEEeqnarray}
\begin{IEEEeqnarray}{l}
0 = G(\mathbf{v}_\text{diode}) + 
\underbrace{\left(\begin{array}{c c c}
        g_{11} + g_{21} + \ldots & 0 & \ldots\\
        0 & g_{12} + g_{22} + \ldots & \ldots\\
        \vdots & \ddots   & \vdots
\end{array}\right)}_{H}\mathbf{v}_\text{diode} +
  \underbrace{\left(\begin{array}{c c c}
        -g_{11} & -g_{21} & \ldots\\
        -g_{12} & -g_{22} & \ldots\\
        \vdots & \ddots   & \vdots
\end{array}\right)}_{B}\mathbf{v}_p \label{eq:feedforward_kernel_1} \\
\mathbf{v}_\text{diode} + H^{-1} Y(\mathbf{v}_\text{diode}) = -B
\mathbf{v}_p.\label{eq:feedforward_kernel_2}
\end{IEEEeqnarray}
\end{strip}

Performing the same analysis as in Section~\ref{sec:example1}, we have a matrix
$\tilde H$ given by \eqref{eq:feedforward_kernel_0}.
This can be verified in a similar manner to the proof of
Proposition~\ref{prop:Stieltjes}: setting the input voltages to zero by shorting the
input ports, and the output currents to zero by opening the output ports, for
example, allows one to solve for the bottom right block using KCL.  The other blocks
can be worked out similarly.  

Neglecting the currents at input and output ports, we have the kernel
equation given by \eqref{eq:feedforward_kernel_1}--\eqref{eq:feedforward_kernel_2}.
Since $H^{-1}$ is diagonal, and each entry of $Y$ is a reverse biased diode and is 
therefore unaffected by scalar multiplication, we can
solve this equation directly for $\mathbf{v}_\text{diode}$:
\begin{IEEEeqnarray*}{rCl}
\mathbf{v}_\text{diode} &=& (I + Y(\mathbf{v}_\text{diode}))^{-1}(-B \mathbf{v}_p)\\
                        &=& \relu{-B\mathbf{v}_p}.\IEEEyesnumber\label{eq:ff_relu}
\end{IEEEeqnarray*}
The output equation is simply $\mathbf{v}_q = \mathbf{v}_\text{diode}$, as we are
neglecting the currents.
The voltage to voltage mapping of this circuit is therefore equal to one
layer of a ReLU neural network with full, entrywise negative weights matrix $-B$.
To remove the restriction of negative weights, we duplicate the input vector of
each layer and negate one of the copies using a negative synaptic weight vector.
This gives a layer of the form
\begin{IEEEeqnarray*}{rCl}
\mathbf{v}_\text{diode} &=& \relu{(-B_1, -B_2)\begin{pmatrix}\boldsymbol{\sigma}_1
\mathbf{v}_p\\\boldsymbol{\sigma}_2 \mathbf{v}_p\end{pmatrix}}\\
                        &=& \relu{(-\boldsymbol{\sigma}_1 B_1 -\boldsymbol{\sigma}_2 B_2)\mathbf{v}_p},\IEEEyesnumber\label{eq:ff_layer}
\end{IEEEeqnarray*}
The matrix $-\boldsymbol{\sigma}_1 B_1 -\boldsymbol{\sigma}_2 B_2$ is now arbitrary.  By cascading such layers,
we can reproduce any ReLU neural network.  Furthermore, by including
crossbar arrays with the port configuration of Section~\ref{sec:example1}, we can
implement ReLU neural networks with equilibrium layers.

\section{Other activations}\label{sec:activations}

So far, we have only considered ideal diodes and their computational counterpart, the
ReLU.  In this section, we first examine the non-ideal diode, and then give a circuit
implementation for a saturation activation function.  

\subsection{Non-ideal diodes and the ``diode ReLU''}

\begin{figure}[hb]
    \centering
    \includegraphics[width=0.7\linewidth]{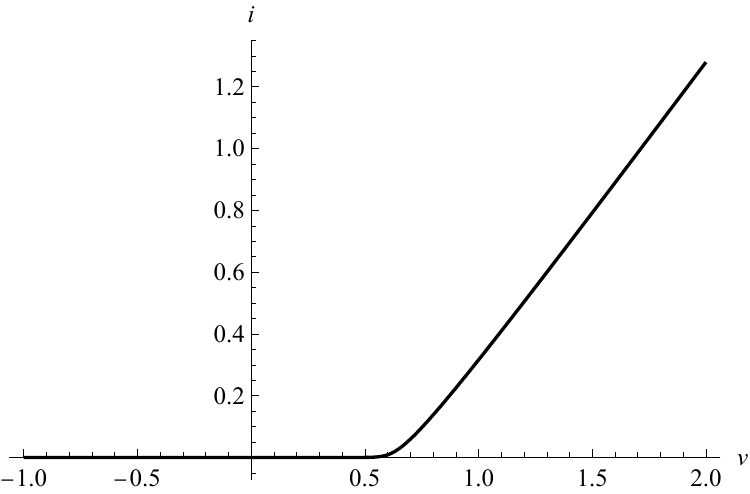}
    \caption{The diode ReLU, plotted for $v_T = 25.852$ mV, $n=1$ and $i_s = 1$ pA.}
    \label{fig:diode_ReLU}
\end{figure}

A more realistic model of a forward biased diode is given by the \emph{Shockley equation}:
\begin{IEEEeqnarray}{rCl}
    v = Z_S(i) := n v_T \log\left(\frac{i}{i_s} + 1\right),
\end{IEEEeqnarray}
where $n$, $v_T$ and $i_s$ are physical constants. \blue{In the ideal case, the
    resolvent of the diode gives the ReLU activation.  To find the activation
corresponding to the Shockley equation, we again}
compute the resolvent, $(Z_S + I)^{-1}$, \blue{by inverting the map $i \mapsto Z_S(i)
+ i$.  This gives}
\begin{IEEEeqnarray}{rCl}
    i =(Z_S + I)^{-1}(v) = n v_T W\left(\frac{i_s\exp\left(\frac{v + i_s}{nv_T}\right)}{n v_T}\right) - i_s,
\end{IEEEeqnarray}
where $W$ is the Lambert function, defined as the inverse of the function $V(x) =
xe^x$, which has a unique solution when $x > 0$.  This equation defines a smoothed
ReLU, plotted in Figure~\ref{fig:diode_ReLU}, which we call the \emph{diode ReLU}.
As in the case of the ideal diode, a dual diode ReLU mapping current to voltage can be
implemented using the admittance of a reverse biased diode.

Replacing the ideal diode model with the Shockley equation then gives a realization
theorem analogous to Corollary~\ref{cor:correspondence}, but for monDEQs with diode ReLU
activations.  Beyond changing the activation
function, the remaining theory of the paper is unchanged.

\subsection{Saturation}

In this section, we show that a saturation activation can be implemented using ideal
Zener diodes.
The \emph{(forward biased) ideal unit Zener diode} is defined as the circuit element
with \blue{admittance}

\begin{minipage}{0.5\linewidth}
    \blue{
\begin{IEEEeqnarray}{rCl}
i &:=& \begin{cases} (-\infty, 0] & v =-1\\
        0 & -1 < v < 0 \\
        [0, \infty) & v = 0.
      \end{cases}\nonumber\\*
&&\nonumber
\end{IEEEeqnarray}
}
\end{minipage}%
\begin{minipage}{0.5\linewidth}
    \centering
    \includegraphics{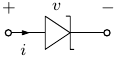}
\end{minipage}
   \vspace{0.2cm}

\noindent The electrical dual of the ideal unit Zener diode is the \emph{ideal unit current
regulator diode (CRD)}, defined as the circuit element with \blue{impedance}

\begin{minipage}{0.5\linewidth}
    \blue{
\begin{IEEEeqnarray}{rCl}
v &:=& \begin{cases} (-\infty, 0] & i =-1\\
        0 & -1 < i < 0 \\
        [0, \infty) & i = 0.
      \end{cases}\nonumber\\*
&&\nonumber
\end{IEEEeqnarray}
}
\end{minipage}%
\begin{minipage}{0.5\linewidth}
    \centering
       \includegraphics{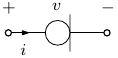}
   \end{minipage}

\noindent An odd current/voltage characteristic can be obtained by connecting a forward biased
Zener diode in series with a reverse biased Zener diode, to give the
\blue{admittance} 

\begin{minipage}{0.5\linewidth}
    \blue{
\begin{IEEEeqnarray}{rCl}
Y_\text{sat}(v) &:=& \begin{cases} (-\infty, 0] & v =-1\\
        0 & -1 < v < 1 \\
        [0, \infty) & v = 1.
      \end{cases}\nonumber\\*
&&\nonumber
\end{IEEEeqnarray}
}
\end{minipage}%
\begin{minipage}{0.5\linewidth}
    \centering
       \includegraphics{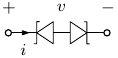}
\end{minipage}

\begin{figure}[hb]
    \centering
    \includegraphics{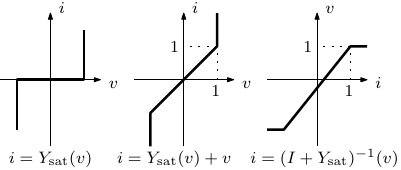}
    \caption{The saturation activation may be realized as the resolvent of two
    opposing Zener diodes, which give the admittance $Y_\text{sat}$.}
    \label{fig:sat_resolvent}
\end{figure}

\noindent Taking the resolvent of this admittance then gives the saturation,
\begin{IEEEeqnarray}{rCl}
v = (I + Y_\text{sat})^{-1}(i) &=& \begin{cases} 1 & i > 1\\
        i & -1 \leq i \leq 1 \\
        -1 & i < -1,
      \end{cases}
\end{IEEEeqnarray}
\blue{as illustrated in Figure~\ref{fig:sat_resolvent}.}

\noindent The electrical dual of the opposing pair of Zener diodes is the parallel interconnection of a forward
biased CRD and a reverse biased CRD, giving the \blue{impedance}

\begin{minipage}{0.5\linewidth}
    \blue{
\begin{IEEEeqnarray}{rCl}
Z_\text{sat}(i) &:=& \begin{cases} (-\infty, 0] & i =-1\\
        0 & -1 < i < 1 \\
        [0, \infty) & i = 1.
      \end{cases}\nonumber\\*
&&\nonumber
\end{IEEEeqnarray}
}
\end{minipage}%
\begin{minipage}{0.5\linewidth}
    \centering
       \includegraphics{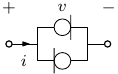}
   \end{minipage}

\noindent Replacing the forward biased diodes in Theorem~\ref{thm:circuit} and
Corollary~\ref{cor:correspondence} with opposing Zener diodes, and the reverse biased
diodes with anti-parallel CRDs,
therefore gives a realization of saturation monDEQs in analog hardware.

\section{Computational experiment}\label{sec:experiment}

In this section, we test the efficacy of hardware linearization using a
hardware-level simulation of a simple crossbar cascade network, and compare its
performance to the equilibrium backpropagation method of \cite{Winston2020}.

The
network consists of four crossbar arrays, configured as in Figure~\ref{fig:cascade},
with an additional set of synaptic weights at the input ports of the first network.
The numbers of ports between each layer are $(10, 9, 7, 8, 4)$.  The training parameters are
the resistances, current offsets at the output ports of each layer and synaptic
weights.  A random instantiation of the circuit is used to generate 10 training
samples, which are then used to train the network with initial resistances of
$100\,\Omega$, current offsets of zero and synaptic weights equal to the layer index
(1 to 4).

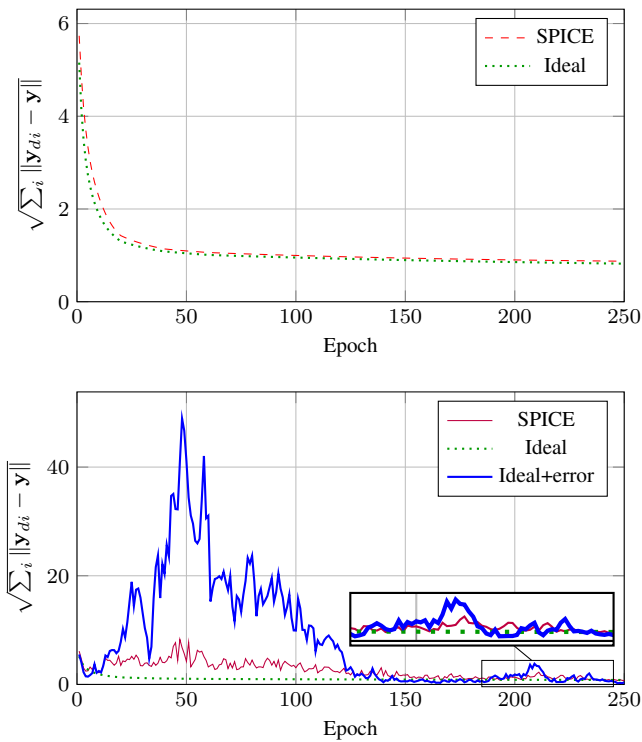
\begin{figure}[h!]
    \footnotesize
    \begin{tikzpicture}[spy using outlines={rectangle, magnification=2, connect spies}]
  \begin{groupplot}[
    group style={
      group size=1 by 2,      
      vertical sep=1.2cm      
    },
    width=\linewidth,
    height=0.618\linewidth,
    xlabel={Epoch},
    ylabel={$\sqrt{\sum_i\norm{\by_{di} - \by}^2}$},
    grid=both,
    xmin=0, xmax=250,
    ymin=0,
    legend style={anchor=north,legend columns=1},
    legend pos=north east
  ]

    \nextgroupplot
    \addplot[red, dashed, mark=none] 
      table [x expr=\coordindex+1, y=errors, col sep=comma] {errors_SPICE_nonideal_0_error.csv};
    \addlegendentry{SPICE}

    \addplot[green!60!black, thick, dotted, mark=none] 
      table [x expr=\coordindex+1, y=errors, col sep=comma] {errors_split_ideal_0_error.csv};
    \addlegendentry{Ideal}

    \nextgroupplot
    \addplot[purple, mark=none] 
      table [x expr=\coordindex+1, y=errors, col sep=comma] {errors_SPICE_error_with_comp.csv};
    \addlegendentry{SPICE}

    \addplot[green!60!black, thick, dotted, mark=none] 
      table [x expr=\coordindex+1, y=errors, col sep=comma] {errors_split_ideal_0_error.csv};
    \addlegendentry{Ideal}

    \addplot[blue, thick, mark=none] 
      table [x expr=\coordindex+1, y=errors, col sep=comma] {errors_split_error_no_comp.csv};
    \addlegendentry{Ideal+error}

        \coordinate (spypoint) at (axis cs:215,2); 
    \coordinate (spyinset) at (185, 12); 
    \spy[width=3.5cm, height=0.7cm] on (spypoint) in node[fill=white] at (spyinset);

  \end{groupplot}
\end{tikzpicture}
\caption{Training curves for the experiment of Section~\ref{sec:experiment}.  Top:
Training curves with gradients computed using a SPICE simulation of hardware
linearization, with realistic diode models, and using an idealised model, showing almost
identical performance.  Bottom: training curves with error introduced in both the
nominal device values and the gradient updates.  A SPICE simulation of hardware
linearization is compared with gradients computed using an idealised model which is
agnostic to the errors.  The ideal curve from the top plot is included for reference.}
\label{fig:experiment}
\end{figure}

Stochastic gradient descent is performed on the objective function $\norm{\by -
\by_d}^2$, using several methods to compute the gradient.
Figure~\ref{fig:experiment} (top) shows training curves using equilibrium
backpropagation on an idealized model and using hardware linearization simulated
using the ngSPICE-41 circuit simulator.  Diodes in the circuit simulation are
modelled with a breakdown voltage of $50$ V, a reverse-bias saturation current of
$1\times10^{-13}$ A and an ideality factor of $1.05$.  \blue{For the purposes of
    hardware linearization, the diode is considered to be conducting if $v > 0$ and
non-conducting if $v \leq 0$}.  The training curves are almost
identical, indicating that, although hardware linearization assumes ideal diodes, the
error introduced by realistic diode models does not have a significant detrimental
effect on the training performance.

The second experiment, illustrated in the bottom of Figure~\ref{fig:experiment},
investigates the robustness of training to errors both in the nominal resistances and
in the gradient updates, simulating the effect of switching errors in memristors.
Initial resistances are sampled from a normal distribution, centered at $100\,
\Omega$, with a standard deviation of $5\%$.  At each training step, the updated resistance
value is normally distributed about its nominal value with a standard deviation of $10\%$.  
Training is compared using hardware linearization on the true circuit model, which
automatically takes into account errors from nominal resistances, and equilibrium
backpropagation performed on the nominal model.  It can be seen that the training
curve for the latter technique exhibits much larger transients in error than that of
the former.  It is also interesting to note that both methods eventually outperform
the ideal, noise-free training procedure of the previous experiment.  This is in line
with previous observations that adding noise to the gradient improves learning
\cite{Neelakantan2015}.

\blue{A more realistic simulation would model the effects of parasitic reactance.  
    Such effects will have no impact on the stability of the
    circuit: the circuit will remain a feedforward interconnection of passive
    circuits \cite{Hughes2017} and, subject to non-zero resistance on
    each branch, which is generically the case in a non-ideal realization, the
    circuit will also be stable.  Parasitic reactance may, however, limit the
    frequency at which hardware linearization may be performed, and a hardware
    realization that mitigates this effect is an interesting topic for future
research.}

%
%
%
%
%
%
%

\section{Conclusions}\label{sec:outlook}

Corollary~\ref{cor:correspondence} establishes a correspondence between
certain passive nonlinear networks and a class of equilibrium neural networks.  This gives a
theoretical basis for the analysis and design of analog machine learning hardware.
We have furthermore shown that such hardware can be trained using hardware
measurements of the gradient, by \emph{hardware linearization}, introduced in
Theorems~\ref{thm:hardware_linearization} and~\ref{thm:reciprocal_backprop}.
As a running example, we have characterized the circuits that can be realized using
crossbar arrays, a common configuration of adjustable resistances used in machine
learning multipliers.  While using a single crossbar array allows only a restricted
class of networks to be realized, cascading crossbar arrays with adjustable
\emph{synaptic} gains allows a class of networks with universal approximation
properties to be realized.  The conductances of the array and the synaptic gains are
in direct analogy with the conductances and synaptic weights of conductance-based
models of biological nervous systems.

The focus of the paper has been on ReLU activations implemented with ideal diodes,
however we have also given realizations of saturation activations, and
shown that non-ideal diodes give rise to a novel activation, the
\emph{diode ReLU}.  A basic device-level simulation demonstrates that hardware
linearization is effective in practice and is robust to device nonidealities.

A first direction for future work is to demonstrate hardware linearization in machine
learning applications.  This requires the development of a circuit simulation tool
which integrates with machine learning frameworks; while accurate, existing circuit
simulators do not scale well enough for learning applications, as also noted in
\cite{Kendall2020, Scellier2024}.

A second interesting line of investigation is the realization of dynamic
neural networks, such as Recurrent Equilibrium Networks \cite{Revay2023,
Martinelli2023}, as nonlinear RLC networks.  This would allow the framework we
present here to be extended to machine learning circuits with reactive elements, such
as Hopfield neural networks \cite{Hopfield1985} and KirchhoffNets \cite{Gao2025}.

Finally, generalizing the results of this paper from Euclidean monotonicity to more
general definitions of monotonicity \cite{Davydov2024} is an interesting avenue for
future research.

\section{Acknowledgement}
This research has been shaped by conversations with many colleagues, among them
Sebastian Banert, Bart Besselink, Thiago Burghi, Pontus Giselsson,
Fulvio Forni, Guillaume Hennequin, Anne-Men Huijzer, Angus O'Grady, Richard Pates,
Benjamin Scellier, Arjan van der Schaft, Rodolphe Sepulchre, Malcolm Smith and Henk van Waarde.  All
responsibility rests with the author.

    \section*{Appendix: Auxiliary results}
    \begin{lemma}\label{lem:inverse_derivative}
        Given a map $K: \R \to \R^{n\times n}$, if
        $K(\theta)$ is non-singular at $\theta$, then 
        \begin{IEEEeqnarray}{rCl}
        \td{}{\theta} K(\theta)^{-1} &=& - K(\theta)^{-1}\td{K(\theta)}{\theta}
        K(\theta)^{-1}.
        \end{IEEEeqnarray}
    \end{lemma}

    \begin{proof}
        \begin{IEEEeqnarray*}{+l+x*}
        K(\theta)K(\theta)^{-1} = I\\
        \td{}{\theta}\left(K(\theta)K(\theta)^{-1}\right) = 0\\
        \td{K(\theta)}{\theta} K(\theta)^{-1} + K(\theta) \td{}{\theta} K(\theta)^{-1}
                                                          = 0\\
        \td{}{\theta}K(\theta)^{-1} = -K(\theta)^{-1} \td{K(\theta)}{\theta}
        K(\theta)^{-1}.& \qedhere
        \end{IEEEeqnarray*}
    \end{proof}

\begin{lemma}
    \blue{
\label{lem:implicit_diff}
Let $\Theta \subseteq \R^r$ be an open set, let $\theta \mapsto H_\theta,\
B_\theta,\ \bu_\theta$ be $C^1$ on $\Theta$, let $\psi : \R^n \rightrightarrows
\R^n$ be a set-valued mapping, and suppose that $H_\theta \succ 0$ for all $\theta$
and no diode is at its switching point. Fix $\bar\theta \in \Theta$ and write $\bz := \bz(\bar\theta)$ and
$\boldsymbol{\eta} := -H_{\bar\theta}\,\bz - B_{\bar\theta}\,\bu_{\bar\theta}
\in \psi(\bz)$. Then $\nabla_\theta \bz$, evaluated at $\bar\theta$, satisfies
\begin{IEEEeqnarray}{rCl}
0 &\in& H_{\bar\theta}\,\nabla_\theta \bz
   + D\psi(\bz \mid \boldsymbol{\eta})\bigl(\nabla_\theta \bz\bigr)
   + B_{\bar\theta}\,\nabla_\theta \bu_\theta \nonumber\\
&& \negthickspace{}+\ \nabla_\theta H_\theta\,\bz
   + \nabla_\theta B_\theta\,\bu_\theta. \IEEEyesnumber\label{eq:implicit_diff_lemma}
\end{IEEEeqnarray}
}
\end{lemma}

\begin{proof}
    We begin by showing that on $\Theta$ there is a
single-valued $C^1$ branch $\theta \mapsto \bz(\theta)$ of solutions to
\begin{IEEEeqnarray}{rCl}
0 &\in& H_\theta\, \bz(\theta) + \psi\bigl(\bz(\theta)\bigr)
        + B_\theta\, \bu_\theta. \IEEEyesnumber\label{eq:eq_branch}
\end{IEEEeqnarray}
The parameter space $\Theta$ may be partitioned into open sets on which the diodes are either
conducting or non-conducting, separated by a switching set on which one or more diode
is at the switching point.  Denote the open set on which the circuit is operating by
$\Theta_0$.  On $\Theta_0$, a subset of the diode variables are zero (voltages over
conducting forward biased diodes, currents over non-conducting reverse biased
diodes).  The remaining variables, denote them $\bz_0$, satisfy the linear equation
\begin{IEEEeqnarray}{rCl}
0 = H^0_\theta \bz_0 + B^0_\theta \bu_\theta,\label{eq:linear_z}
\end{IEEEeqnarray}
in a neighbourhood of $\bar\theta$,
where $H^0_\theta$ and $B^0_\theta$ are the appropriate submatrices of $H_\theta$ and
$B_\theta$.  Since $H_\theta$ and $B_\theta$ are assumed to be $C^1$ in $\theta$ and
$H_\theta$ is assumed to be invertible,
\eqref{eq:linear_z} defines a $C^1$ mapping from $\theta$ to $\bz_0(\theta)$, and
hence $\bz(\theta)$, in a neighbourhood of $\bar\theta$.

We now rewrite \eqref{eq:implicit_diff_lemma} as a tangent cone inclusion. 
By \eqref{eq:eq_branch}, the $C^1$ selection
\begin{IEEEeqnarray}{rCl}
\boldsymbol{\eta}(\theta) &:=& -H_\theta\,\bz(\theta) - B_\theta\,\bu_\theta
\IEEEyesnumber\label{eq:eta_def}
\end{IEEEeqnarray}
satisfies $\boldsymbol{\eta}(\theta) \in \psi(\bz(\theta))$ in a neighbourhood of
$\bar\theta$, with $\boldsymbol{\eta}(\bar\theta) = \boldsymbol{\eta}$.
Differentiating \eqref{eq:eta_def} at $\bar\theta$ gives
\begin{IEEEeqnarray*}{rCl}
\nabla_\theta \boldsymbol{\eta}
&=& -H_{\bar\theta}\,\nabla_\theta \bz
    - \nabla_\theta H_\theta\,\bz
    - B_{\bar\theta}\,\nabla_\theta \bu_\theta
    - \nabla_\theta B_\theta\,\bu_\theta,\\
    \IEEEyesnumber\label{eq:eta_derivative}
\end{IEEEeqnarray*}
and substituting \eqref{eq:eta_derivative} into \eqref{eq:implicit_diff_lemma}
reduces the lemma to the inclusion
\begin{IEEEeqnarray}{rCl}
\nabla_\theta \boldsymbol{\eta}
&\in& D\psi(\bz \mid \boldsymbol{\eta})(\nabla_\theta \bz),
    \IEEEyesnumber\label{eq:chain_rule_form}
\end{IEEEeqnarray}
which by definition of the graphical derivative is equivalent to
\begin{IEEEeqnarray}{rCl}
\bigl(\nabla_\theta \bz \cdot u,\, \nabla_\theta \boldsymbol{\eta} \cdot u\bigr)
&\in& T_{\operatorname{gph}\psi}(\bz, \boldsymbol{\eta})
\quad\text{for all } u \in \R^r.
    \IEEEyesnumber\label{eq:tangent_form}
\end{IEEEeqnarray}

We now verify \eqref{eq:tangent_form} directly.
Fix $u \in \R^r$ and consider the $C^1$ curve
\begin{IEEEeqnarray*}{rCl}
\gamma_u(t) &:=& \bigl(\bz(\bar\theta + t u),\,
\boldsymbol{\eta}(\bar\theta + t u)\bigr),
\end{IEEEeqnarray*}
defined for $|t|$ small enough that $\bar\theta + t u \in \Theta$. From above, $\gamma_u(t) \in \operatorname{gph}\psi$ for all such
$t$, while $\gamma_u(0) = (\bz, \boldsymbol{\eta})$ and $\gamma_u'(0) =
(\nabla_\theta \bz \cdot u, \nabla_\theta \boldsymbol{\eta} \cdot u)$ by the
classical chain rule.
Choose any sequence $\tau^\nu \downarrow 0$ small enough that $\gamma_u(\tau^\nu)$
is defined, and set
$x^\nu := \gamma_u(\tau^\nu) \in \operatorname{gph}\psi.$
Then $x^\nu \to (\bz, \boldsymbol{\eta})$ by continuity of $\gamma_u$, and
\begin{IEEEeqnarray*}{rCl}
\frac{x^\nu - (\bz, \boldsymbol{\eta})}{\tau^\nu}
&=& \frac{\gamma_u(\tau^\nu) - \gamma_u(0)}{\tau^\nu}
\ \to\ \gamma_u'(0)
\end{IEEEeqnarray*}
by the definition of $\gamma_u'(0)$. Equation \eqref{eq:tangent_form} then follows
directly from the definition of the tangent cone.
\end{proof}

\printbibliography

\begin{IEEEbiography}[{\includegraphics[width=1in, height=1.25in, clip, keepaspectratio]{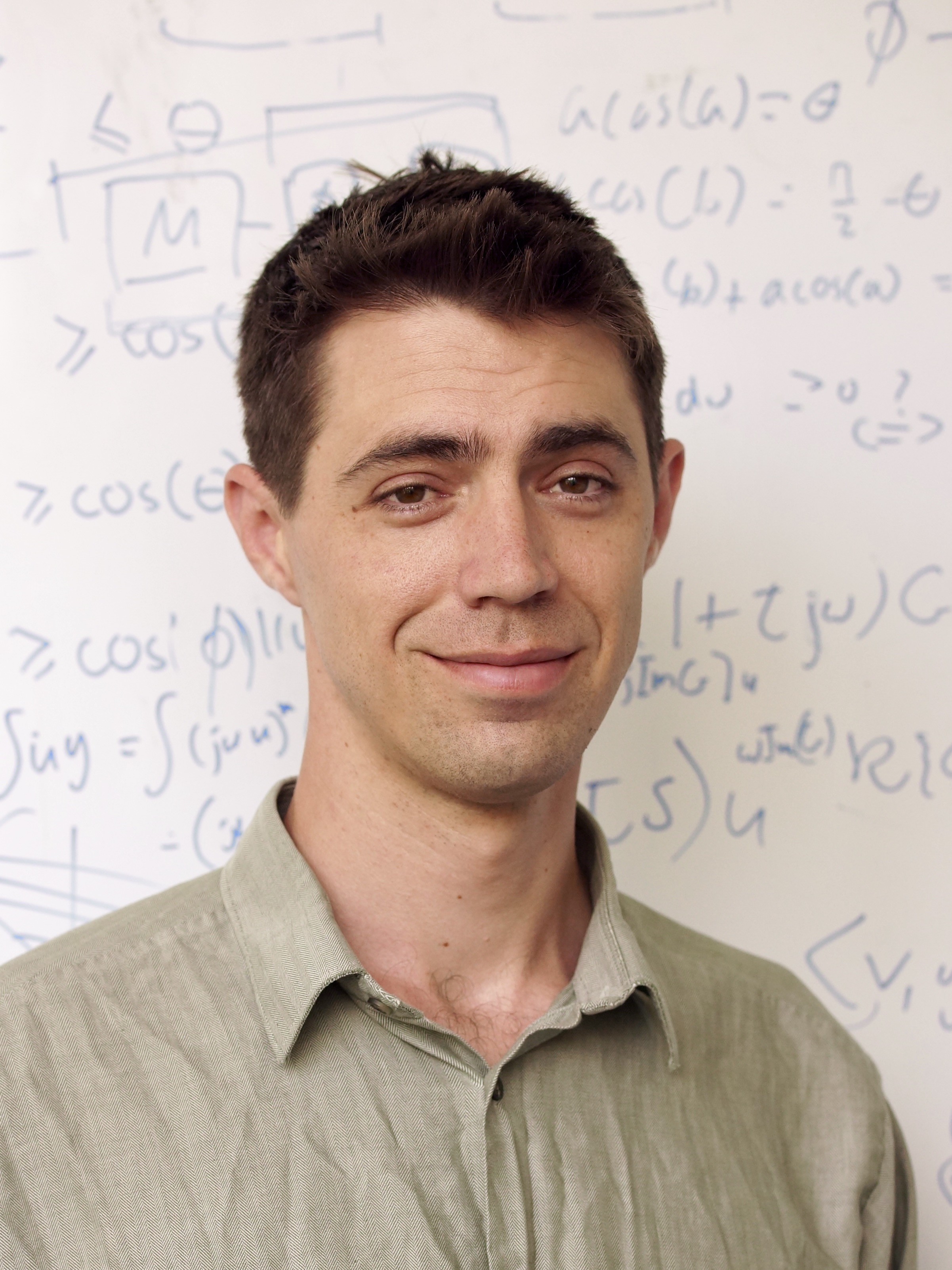}}]
    {Thomas Chaffey} is a lecturer in the School of Electrical and Computer Engineering at the University of Sydney, Australia. He received the B.Sc. (advmath) degree in mathematics and computer science and the M.P.E. degree in mechanical engineering from the University of Sydney in 2015 and 2018, respectively, and the Ph.D. degree from the University of Cambridge, U.K., in 2022.  From 2022 to 2025 he held the Maudslay-Butler Research Fellowship in Engineering at Pembroke College, University of Cambridge. His research interests are in nonlinear control and its intersection with optimization, circuit theory and learning.
\end{IEEEbiography}
\end{document}